\documentclass{jpp}
\usepackage{graphicx}
\usepackage{natbib}
\usepackage{amsmath}
\usepackage{amssymb}
\usepackage{color}
\usepackage{authblk}
\usepackage{media9}
\usepackage{appendix}

\pdfoutput=1

\ifCUPmtlplainloaded \else
  \checkfont{eurm10}
  \iffontfound
    \IfFileExists{upmath.sty}
      {\typeout{^^JFound AMS Euler Roman fonts on the system,
                   using the 'upmath' package.^^J}%
       \usepackage{upmath}}
      {\typeout{^^JFound AMS Euler Roman fonts on the system, but you
                   dont seem to have the}%
       \typeout{'upmath' package installed. JPP.cls can take advantage
                 of these fonts, if you use 'upmath' package.^^J}%
      }
  \else
  \fi
\fi

\ifCUPmtlplainloaded \else
  \checkfont{msam10}
  \iffontfound
    \IfFileExists{amssymb.sty}
      {\typeout{^^JFound AMS Symbol fonts on the system, using the
                'amssymb' package.^^J}%
       \usepackage{amssymb}%
       \let\le=\leqslant  
       \let\ge=\geqslant  
      }{}
  \fi
\fi

\ifCUPmtlplainloaded \else
  \checkfont{msam10}
  \iffontfound
    \IfFileExists{amssymb.sty}
      {\typeout{^^JFound AMS Symbol fonts on the system, using the
                'amssymb' package.^^J}%
       \usepackage{amssymb}%
       \let\le=\leqslant  
       \let\ge=\geqslant  
      }{}
  \fi
\fi

\ifCUPmtlplainloaded \else
\IfFileExists{amsbsy.sty}
             {\typeout{^^JFound the 'amsbsy' package on the system, using it.^^J}%
     \usepackage{amsbsy}}
    {}
\fi

\newsavebox{\astrutbox}
\sbox{\astrutbox}{\rule[-5pt]{0pt}{20pt}}

\newcommand{\Alfven}{Alfv\'{e}n }
\newcommand{\Alfvenic}{Alfv\'{e}nic }
\newcommand{\T}[1]{{\tt #1}}
\newcommand{\V}[1]{\mathbf{#1}}
\newcommand{\xhat}{\mbox{$\hat{\mathbf{x}}$}}
\newcommand{\yhat}{\mbox{$\hat{\mathbf{y}}$}}
\newcommand{\zhat}{\mbox{$\hat{\mathbf{z}}$}}

\newcommand{\figref}[1]{Fig.~\ref{#1}}
\newcommand{\secref}[1]{\S\ref{#1}}

\title[Localized \Alfven Wavepacket Collisions]{Nonlinear energy transfer and current sheet development in localized Alfv\'en wavepacket collisions in the strong turbulence limit}
    
\author[1]{J.~L.~Verniero}
\author[2]{G.~G.~Howes}
\author[3]{K.~G.~Klein}

\affil[1]{Department of Mathematics, University of Iowa, Iowa City IA 54224, USA}
\affil[2]{Department of Physics and Astronomy,University of Iowa, Iowa City IA 54224, USA}
\affil[3]{Department of Climate and Space Sciences and Engineering,University of Michigan, Ann Arbor, MI 48109, USA}
\pubyear{2010}
\volume{650}
\pagerange{119--126}
\date{?; revised ?; accepted ?. - To be entered by editorial office}

\begin{document}
\maketitle

\begin{abstract}
In space and astrophysical plasmas, turbulence is responsible for
transferring energy from large scales driven by violent events or
instabilities, to smaller scales where turbulent energy is ultimately converted into
plasma heat by dissipative mechanisms. The nonlinear interaction between
counterpropagating Alfv\'en waves, denoted \Alfven wave collisions, drives this turbulent energy
cascade, as recognized by early work with incompressible
magnetohydrodynamic (MHD) equations. Recent work employing analytical calculations and nonlinear gyrokinetic simulations of \Alfven wave collisions in an idealized periodic initial state have demonstrated the key properties that strong \Alfven wave collisions mediate effectively the transfer of energy to smaller perpendicular scales and self-consistently generate current sheets. For the more realistic case of the collision between two initially separated \Alfven wavepackets, we use a nonlinear gyrokinetic simulation to show here that these key properties persist: strong \Alfven wavepacket collisions indeed facilitate the perpendicular cascade of energy and give rise to current sheets. Furthermore, the evolution shows that nonlinear interactions occur only while the wavepackets overlap, followed by a clean separation of the wavepackets with straight uniform magnetic fields and the cessation of nonlinear evolution in between collisions, even in the gyrokinetic simulation presented here which resolves dispersive and kinetic effects beyond the reach of the MHD theory.
\end{abstract}



\section{Introduction}
Turbulence plays an important role in facilitating particle
energization and plasma heating in space and astrophysical plasmas,
influencing the macroscopic evolution of many poorly understood
systems, such as the solar corona and solar wind, planetary
magnetospheres, and black hole accretion disks. The turbulent cascade
mediates the transfer of energy from magnetic fields and plasma flows
at large scales down to much smaller scales where dissipation mechanisms
can effectively remove energy from the turbulent fluctuations,
ultimately converting that energy to plasma heat.  Understanding the
details of this nonlinear turbulent cascade to small scales and of the
mechanisms by which the turbulent energy is thermalized represents a
grand challenge problem in heliophysics and astrophysics.

Early research on incompressible magnetohydrodynamics (MHD) turbulence in the 1960s
\citep{Iroshnikov:1963,Kraichnan:1965} suggested that nonlinear
interactions between counterpropagating \Alfven waves---or \Alfven
wave collisions---support the turbulent cascade of energy from large
to small scales. Following significant previous studies on weak
incompressible MHD turbulence
\citep{Sridhar:1994,Montgomery:1995,Ng:1996,Galtier:2000}, recent work
has elucidated the mechanism of energy transfer in \Alfven wave
collisions in the weakly nonlinear limit by computing an asymptotic
analytical solution using incompressible MHD \citep{Howes:2013a},
verifying that solution using nonlinear gyrokinetic simulations in the
MHD limit of perpendicular scales larger than the Larmor radius,
$k_\perp \rho_i \gg 1$ \citep{Nielson:2013a}, and confirming the
results experimentally in the laboratory
\citep{Howes:2012b,Howes:2013b,Drake:2013,Drake:2014,Drake:2016}.  The
derivation of the analytical solution was possible using the idealized
initial conditions of two overlapping, perpendicularly polarized
\Alfven waves in a periodic geometry, and solving for the nonlinear
evolution of the system. For initial plane \Alfven waves with
wavevectors $\V{k}_1^+ = k_\perp \xhat - k_\parallel \zhat$ and
$\V{k}_1^- = k_\perp \yhat + k_\parallel \zhat$, the nonlinear energy
transfer is mediated by a nonlinearly generated, purely magnetic mode
with wavevector $\V{k}_2^{(0)}= k_\perp \xhat + k_\perp \yhat$, which
can be interpreted as an oscillating shear in the magnetic field along
which the \Alfven waves propagate \citep{Maron:2001,Howes:2017b}.  The
nonlinear interaction between $\V{k}_1^\pm$ and $\V{k}_2^{(0)}$ yields
a secular transfer of energy from the $\V{k}_1^+$ \Alfven wave to an
\Alfven wave with $\V{k}_3^+ = 2 k_\perp \xhat + k_\perp \yhat -
k_\parallel \zhat$, and from the $\V{k}_1^-$ \Alfven wave to an
\Alfven wave with $\V{k}_3^- = k_\perp \xhat+ 2k_\perp \yhat +
k_\parallel \zhat$.  Since the energy is transferred to an \Alfven wave
with a higher perpendicular wavenumber, $|\V{k}_{3_ \perp}^\pm| >
|\V{k}_{1 \perp}^\pm|$, this interaction represents the fundamental
mechanism by which turbulence transfers energy from larger to smaller
scales.

Another important discovery about plasma turbulence followed from the
finding that the nonlinear evolution of MHD turbulence simulations
leads to the development of intermittent current sheets
\citep{Matthaeus:1980,Meneguzzi:1981}, and that the dissipation of
turbulent energy is found to be largely concentrated in the vicinity
of these intermittent current sheets
\citep{Uritsky:2010,Osman:2011,Zhdankin:2013}.  This finding has
motivated significant recent efforts to seek evidence of the spatial
localization of plasma heating by the dissipation of turbulence in
current sheets through statistical analyses of solar wind observations
\citep{Osman:2011,Borovsky:2011,Osman:2012a,Perri:2012a,Wang:2013,Wu:2013,Osman:2014b}
and numerical simulations
\citep{Wan:2012,Karimabadi:2013,TenBarge:2013a,Wu:2013,Zhdankin:2013}.
Although these works clearly demonstrate a connection between current
sheets and plasma heating, the origin of these current sheets in
plasma turbulence remains unknown: do they represent advected flux
tube boundaries \citep{Borovsky:2008,Borovsky:2010}, or are they
generated dynamically by the turbulence itself
\citep{Boldyrev:2011,Zhdankin:2012}?  A significant breakthrough on
this question was the discovery that \Alfven wave collisions in the
strong turbulence limit naturally generate current sheets
\citep{Howes:2016b}, making a connection for the first time between
the nonlinear mechanism governing the transfer of energy to small
scales and the self-consistent development of intermittent current
sheets.

The analytical solution for energy transfer in weak \Alfven wave
collisions \citep{Howes:2013a} and the simulations showing that strong
\Alfven wave collisions naturally generate current sheets
\citep{Howes:2016b} were based on an idealized initial condition in
which two finite-amplitude, plane \Alfven waves are initially
overlapping in a periodic geometry before they begin to interact
nonlinearly. Here we eliminate the unrealistic aspect of those studies
by simulating the strong nonlinear interactions between two initially
separated \Alfven wavepackets, with the aim to determine whether the
general properties of \Alfven wave collisions found in the idealized
case persist in this more realistic case of colliding wavepackets.
Specifically, we focus here on answering two questions: (i) Do
collisions between \Alfven wavepackets still mediate the transfer of
energy to small perpendicular scales?; and (ii) Do \Alfven wavepacket
collisions in the strongly nonlinear limit still lead to the
development of intermittent current sheets?  A companion paper
\citep{Verniero:2017b} will investigate whether the nonlinearly
generated $(k_x/k_\perp,k_y/k_\perp,k_z/k_\parallel)=(1,1,0)$ mode
still mediates the energy transfer in the weakly collisional limit of
\Alfven wavepacket collisions.

In \secref{sec:sim}, we describe the setup of this strong, localized
\Alfven wavepacket collision simulation. The nonlinear evolution of
this simulation is analyzed in \secref{sec:mov}, with particular
emphasis on the perpendicular cascade of energy in \secref{sec:perp},
current sheet development in \secref{sec:cs}, evolution of the energy
in \secref{sec:energy}, and general qualitative properties of
localized \Alfven wavepacket collisions in
\secref{sec:properties}. Conclusions are presented in
\secref{sec:conc}.

\section{Simulation}
\label{sec:sim}
Here we employ the Astrophysical Gyrokinetics code \T{AstroGK}
\citep{Numata:2010} to perform a gyrokinetic simulation of the
nonlinear interaction between two initially separated,
counterpropagating \Alfven wavepackets in the strongly nonlinear
limit.

\T{AstroGK} evolves the perturbed gyroaveraged distribution function
$h_s(x,y,z,\lambda,\varepsilon)$ for each species $s$, the scalar
potential $\varphi$, the parallel vector potential $A_\parallel$, and
the parallel magnetic field perturbation $\delta B_\parallel$
according to the gyrokinetic equation and the gyroaveraged Maxwell's
equations \citep{Frieman:1982,Howes:2006}. Velocity space coordinates
are $\lambda=v_\perp^2/v^2$ and $\varepsilon=v^2/2$. The domain is a
periodic box of size $L_{\perp }^2 \times L_{z}$, elongated along the
straight, uniform mean magnetic field $\V{B}_0=B_0 \zhat$, where all
quantities may be rescaled to any parallel dimension satisfying $L_{z}
/L_{\perp } \gg 1$. Uniform Maxwellian equilibria for ions (protons)
and electrons are chosen, with a realistic mass ratio $m_i/m_e=1836$.
Spatial dimensions $(x,y)$ perpendicular to the mean field are treated
pseudospectrally; an upwind finite-difference scheme is used in the
parallel direction, $z$. Collisions employ a fully conservative,
linearized collision operator with energy diffusion and pitch-angle
scattering \citep{Abel:2008,Barnes:2009}.

\begin{figure}
\centering {\includegraphics[trim=0 85bp 0 85bp,clip,scale = .5]{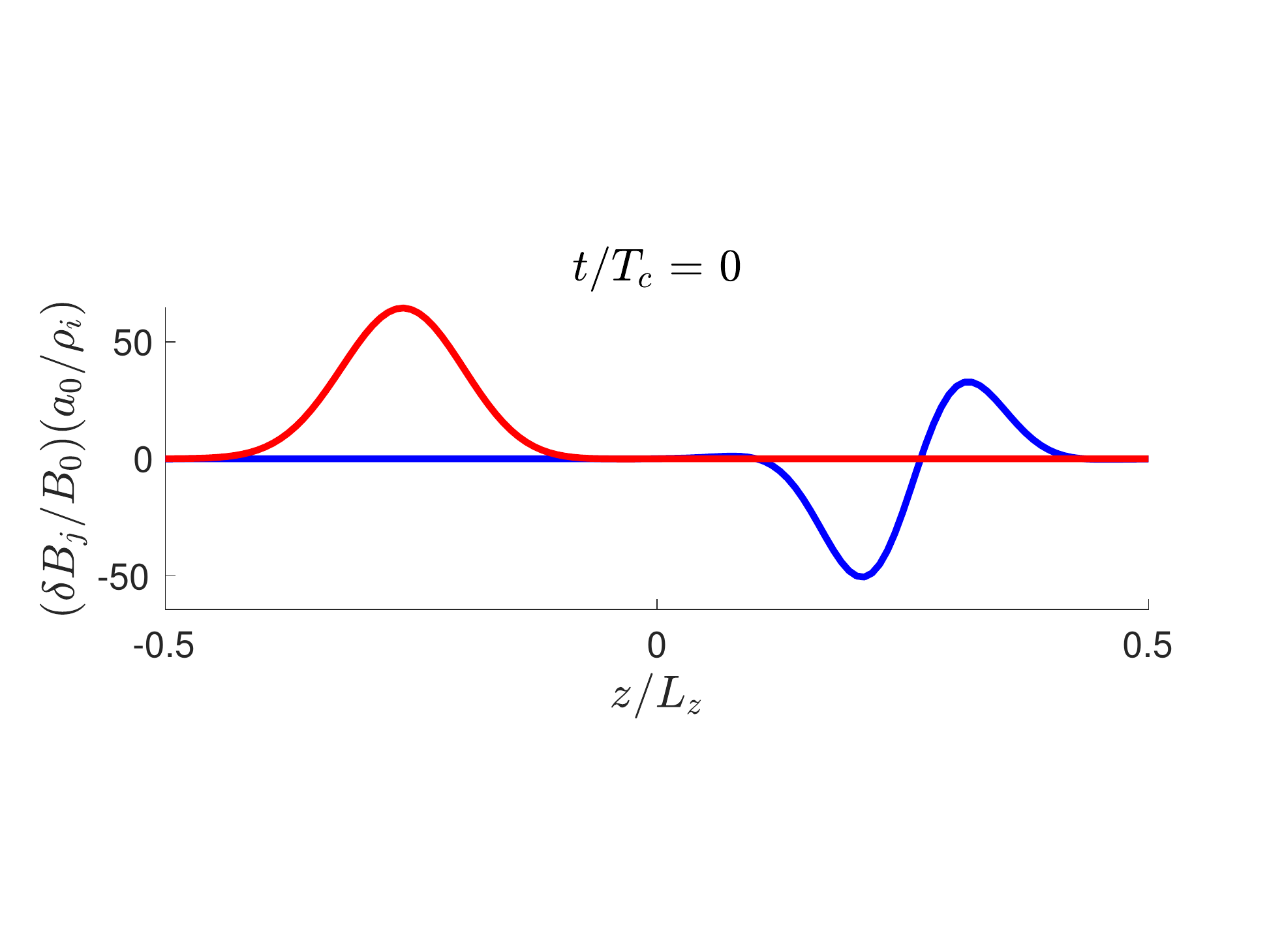}}
\caption{Schematic of the initial conditions specifying the two
  perpendicularly polarized, counterpropagating \Alfven wavepackets
  localized within the periodic domain. Plotted is the $z$-dependence
  of the normalized amplitudes of the perpendicular magnetic field
  perturbation $(\delta B_y/B_0)(a_0/\rho_0)$ for the unipolar
  wavepacket (red) and of the perpendicular magnetic field
  perturbation $(\delta B_x/B_0)(a_0/\rho_0)$ for the dipolar
  wavepacket (blue). The unipolar wavepacket has a perpendicular
  wavevector $\V{k}_\perp^-=(k_x L_\perp,k_y L_\perp)=(1,0)$ and the
  dipolar wavepacket has $\V{k}_\perp^+=(0,1)$.
\label{fig:mysetup}}
\end{figure}

The functional forms along $z$ of the initial \Alfven wavepackets used in this
simulation are shown in \figref{fig:mysetup}. Initially at $z<0$ is a
wavepacket with a magnetic field perturbation $\delta B_y$ polarized
in the $y$ direction, a unipolar variation in $z$ (red), and a
perpendicular structure with wavenumber $\V{k}_\perp^-=(k_x
L_\perp,k_y L_\perp)=(1,0)$. The eigenfunction dictating the different
field components and perturbed distribution functions for this \Alfven
wavepacket is determined by solving the linear, collisionless,
gyrokinetic dispersion relation \citep{Howes:2006} for an \Alfven mode
with the chosen perpendicular Fourier wavevector $(k_x,k_y)$ to obtain
the complex eigenfrequency $\omega$, the complex Fourier coefficients
for the eigenfunctions of the electromagnetic potentials $\hat{\phi}$,
$\hat{A}_\parallel$ and $\delta \hat{B}_\parallel$, and the complex
perturbed gyrokinetic distribution functions for the ions
$\hat{h}_i(v_\parallel,v_\perp)$ and electrons
$\hat{h}_e(v_\parallel,v_\perp)$, where the hat symbol denotes the
$(k_x,k_y)$ Fourier coefficient \citep{Nielson:2013a}. The procedure
for localizing this \Alfven wavepacket in the $z$ direction is
described in Appendix~\ref{appendix:setup}.  Through this procedure,
this unipolar \Alfven wavepacket in \figref{fig:mysetup} propagates in
the $+z$ direction. The other \Alfven wavepacket, initially at $z>0$,
has a magnetic field perturbation $\delta B_x$ polarized in the
$x$ direction, a dipolar structure in $z$ (blue), and a perpendicular
structure with wavenumber $\V{k}_\perp^+=(k_x L_\perp,k_y
L_\perp)=(0,1)$; the eigenfunction specified by the same procedure
dictates that this dipolar wavepacket propagates in the
$-z$ direction. Note that these unsymmetric initial conditions were chosen to further test the complexities in \Alfven wave collisions. In particular by having a unipolar wavepacket, the collision does not shear and unshear in an oscillatory manner as it does in the interaction between two dipolar wavepackets.

The plasma parameters for this strong \Alfven wavepacket collision
simulation are ion plasma beta $\beta_i=1$ and ion-to-electron
temperature ratio $T_i/T_e=1$.  To study the nonlinear evolution of
this \Alfven wavepacket collision in the limit $k_\perp \rho_i\ll 1$,
we choose a perpendicular simulation domain size $L_{\perp}=40 \pi
\rho_i$ with simulation resolution
$(n_x,n_y,n_z,n_\lambda,n_\varepsilon,n_s)= (64,64,128,32,32,2)$.
Therefore, the initial \Alfven wavepackets have perpendicular
wavevectors $\V{k}_\perp^-=(k_x \rho_i,k_y \rho_i)=(0.05,0)$ for the
unipolar wavepacket and $\V{k}_\perp^+=(k_x \rho_i,k_y
\rho_i)=(0,0.05)$ for the dipolar wavepacket, so both waves have the
same initial perpendicular wavenumber $k_\perp^\pm \rho_i=0.05$, but
are polarized perpendicular to each other. The fully resolved
perpendicular range in this dealiased pseudospectral method covers
$0.05 \le k_\perp \rho_i \le 1.05$. Here the ion thermal Larmor radius
is $\rho_i= v_{ti}/\Omega_i$, the ion thermal velocity is $v_{ti}^2 =
2T_i/m_i$, the ion cyclotron frequency is $\Omega_i= q_i B_0/(m_i c)$,
and the temperature is given in energy units.  The parallel length of
the simulation domain is $L_z$, extending over the range
[$-L_z/2$,$L_z/2$].  Note that the simulation domain is triply
periodic, so when a wavepacket exits the domain at $z=\pm L_z/2$, it
re-enters at the opposite end at $z=\mp L_z/2$, enabling these two
wavepackets to undergo successive collisions with each other.  The
linearized Landau collision operator \citep{Abel:2008,Barnes:2009} is
employed with collisional coefficients $\nu_i=\nu_e= 10^{-3}$,
yielding weakly collisional dynamics with $ \nu_s /\omega \ll 1$.

The amplitude of the initial wavepackets is parameterized by the
nonlinearity parameter \citep{Goldreich:1995}, defined by taking the
ratio of the magnitudes of the linear to the nonlinear terms in the
incompressible MHD equations \citep{Howes:2013a,Nielson:2013a}. In
terms of Elsasser variables, defined by $\V{z}^{\pm} = \V{u} \pm
\delta \V{B}/\sqrt{4 \pi (n_{0i} m_i+n_{0e}m_e)}$, the nonlinearity
parameter is defined by \newline
 $\chi^\pm \equiv |\V{z}^{\mp}\cdot \nabla
\V{z}^{\pm}|/|\V{v}_A \cdot \nabla \V{z}^{\pm}|$, where $\chi^\pm$
characterizes the strength of the nonlinear distortion of the
$\V{z}^{\pm}$ \Alfven wave by the counterpropagating $\V{z}^{\mp}$
\Alfven wave. For the particular initial \Alfven wavepackets shown in
\figref{fig:mysetup}, the nonlinearity parameter simplifies to
$\chi^\pm =2 k_\perp \delta B_\perp^\mp /(k_\parallel B_0)$. With the
$\V{z}^{\pm}$ wavepackets having parallel wavenumbers of approximately
$k_\parallel a_0= \mp 3$, where $a_0=L_z/2\pi$, the amplitude of the
unipolar wavepacket $(\delta B_\perp^-/B_0) (a_0/\rho_i) \simeq 60$
gives $\chi^+=2$ and the amplitude of the dipolar wavepacket $(\delta
B_\perp^+/B_0) (a_0/\rho_i) \simeq 40$ gives $\chi^-=1.3$. Strong,
critical balanced turbulence \citep{Goldreich:1995} corresponds to a
nonlinearity parameter of $\chi \sim 1$, so this simulation falls into
the desired limit of strong \Alfven wavepacket collisions.

\section{Evolution of the Nonlinear Interaction}
\label{sec:mov}

\begin{figure}          
\hspace{0.1in} (a) $t/T_c = 0$  \hspace{1.5in}  (b) $t/T_c = 0.5$                        {\includegraphics[scale=.32]{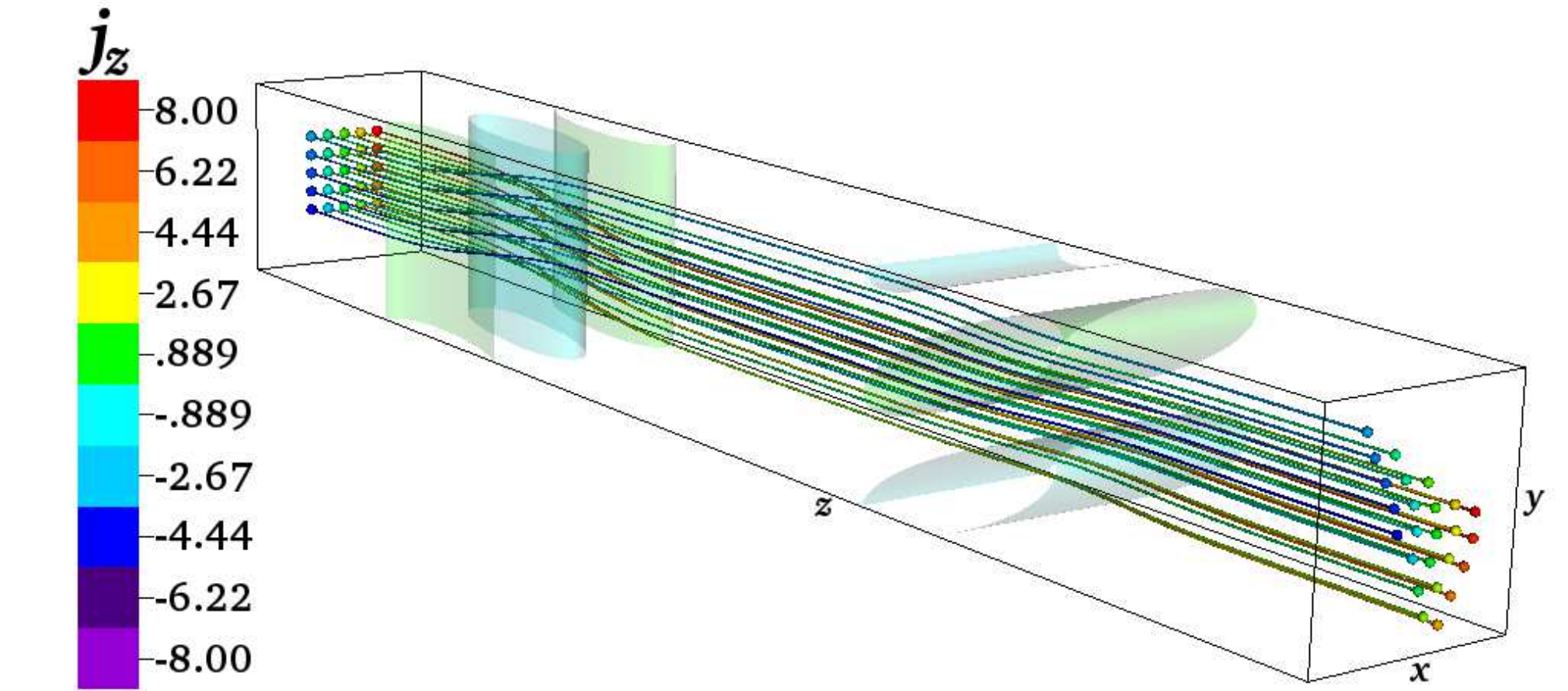}}  \hfill
{\includegraphics[scale=.32]{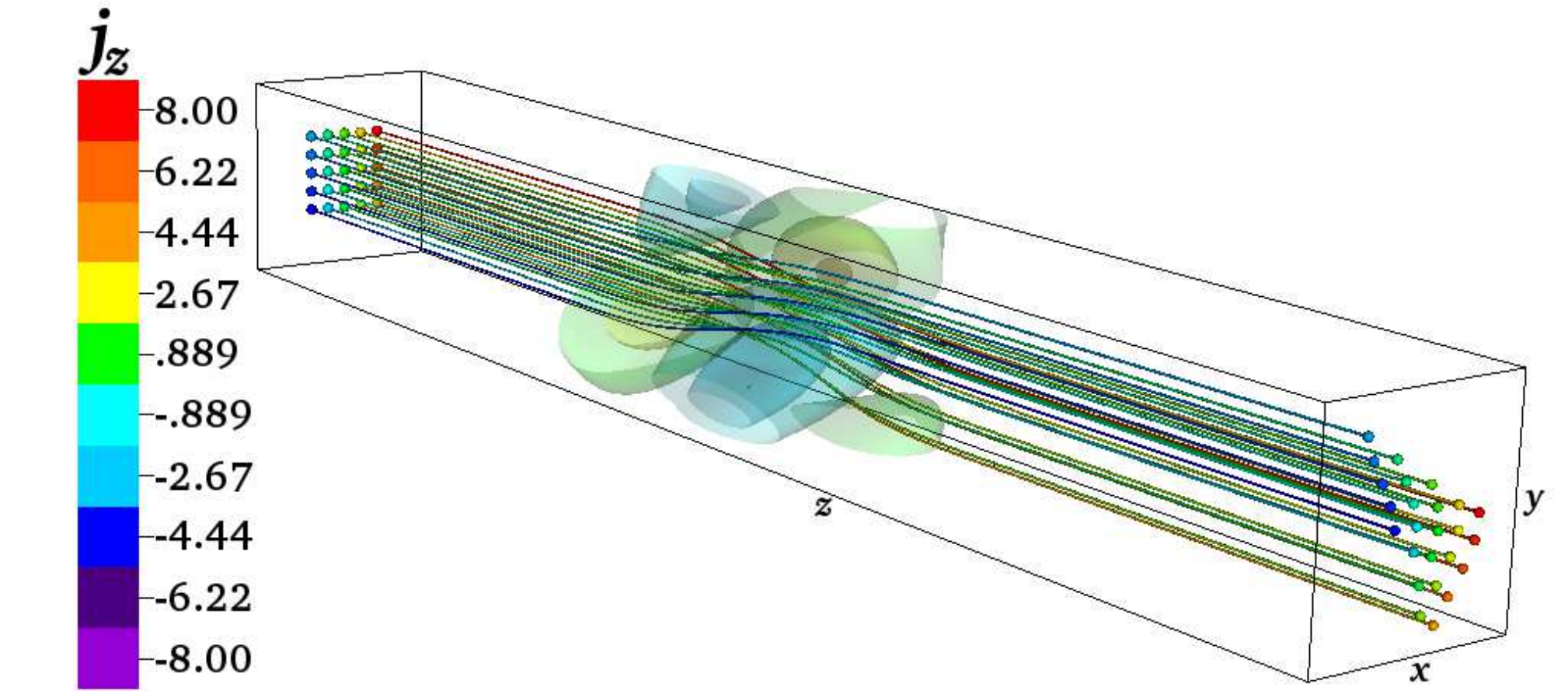}} 
\vfill
\hspace{0.1in} (c) $t/T_c = 1.0$  \hspace{1.5in}  (d) $t/T_c = 1.5$ 
{\includegraphics[scale=.32]{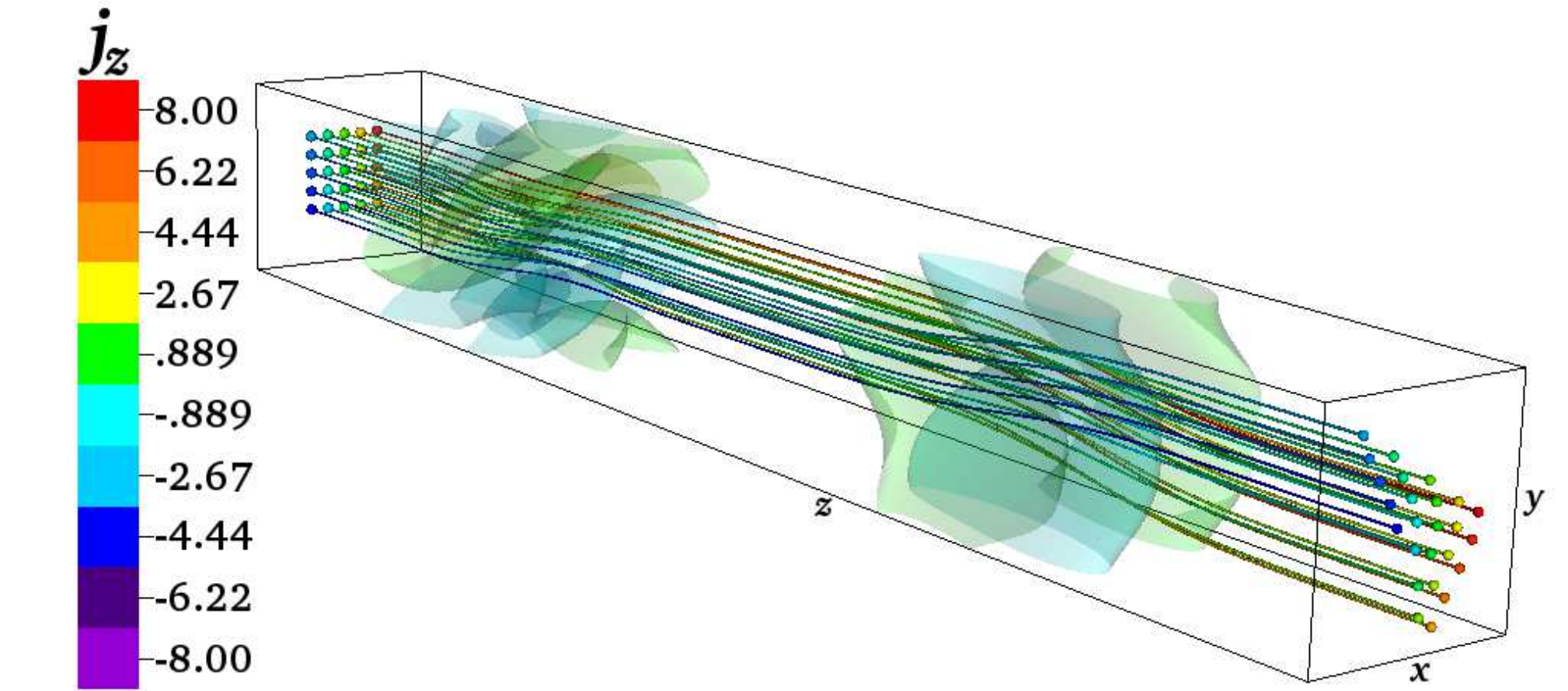}} \hfill  
{\includegraphics[scale=.32]{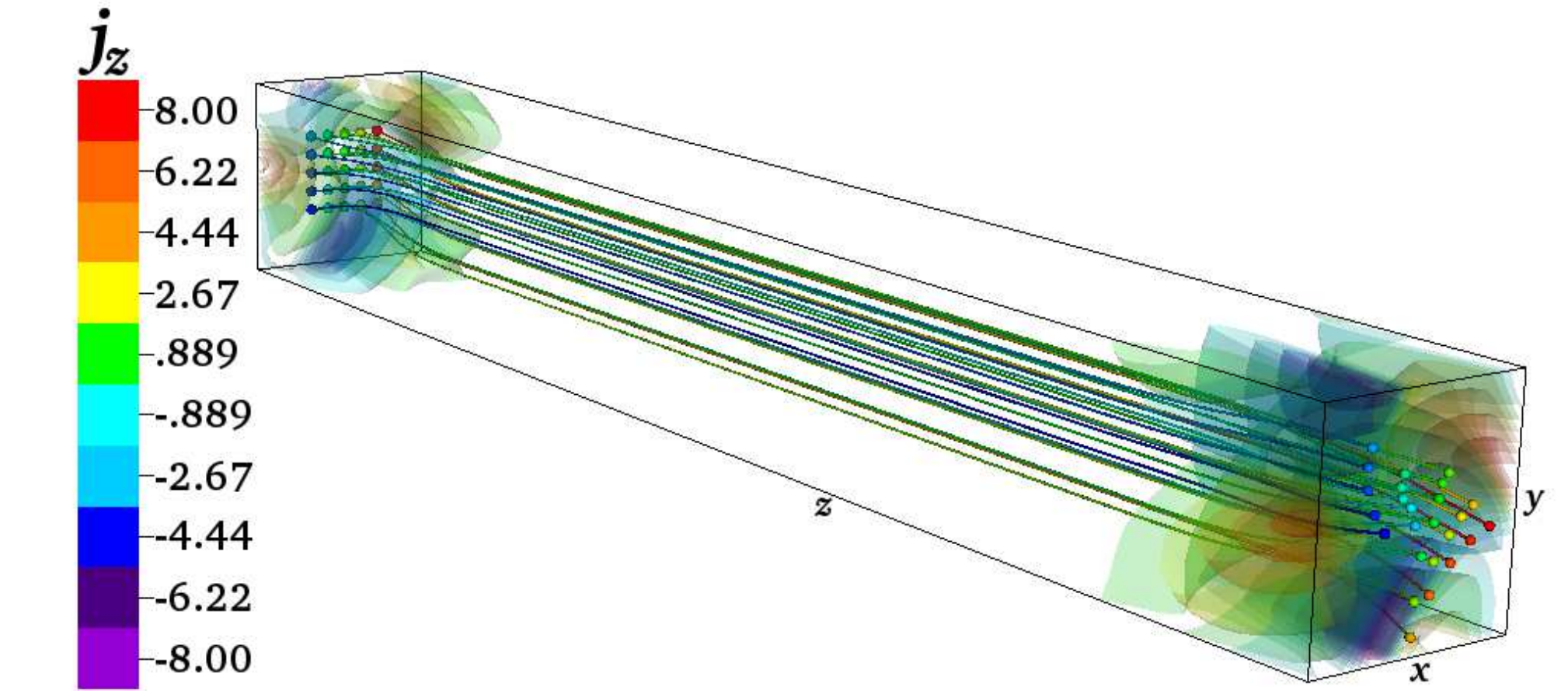}}           
\vfill
\hspace{0.1in} (e) $t/T_c = 2.0$  \hspace{1.5in}  (f) $t/T_c = 2.5$ 
{\includegraphics[scale=.32]{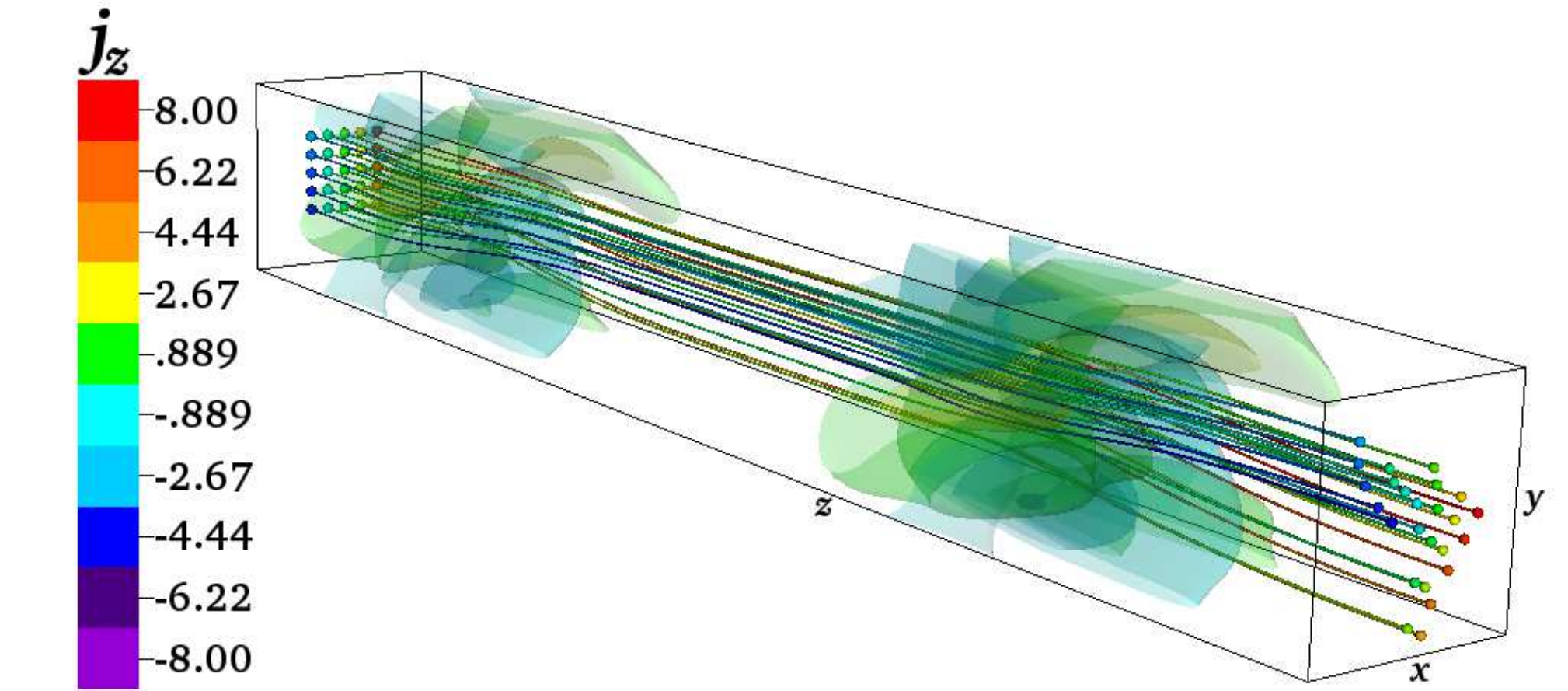}} \hfill  
{\includegraphics[scale=.32]{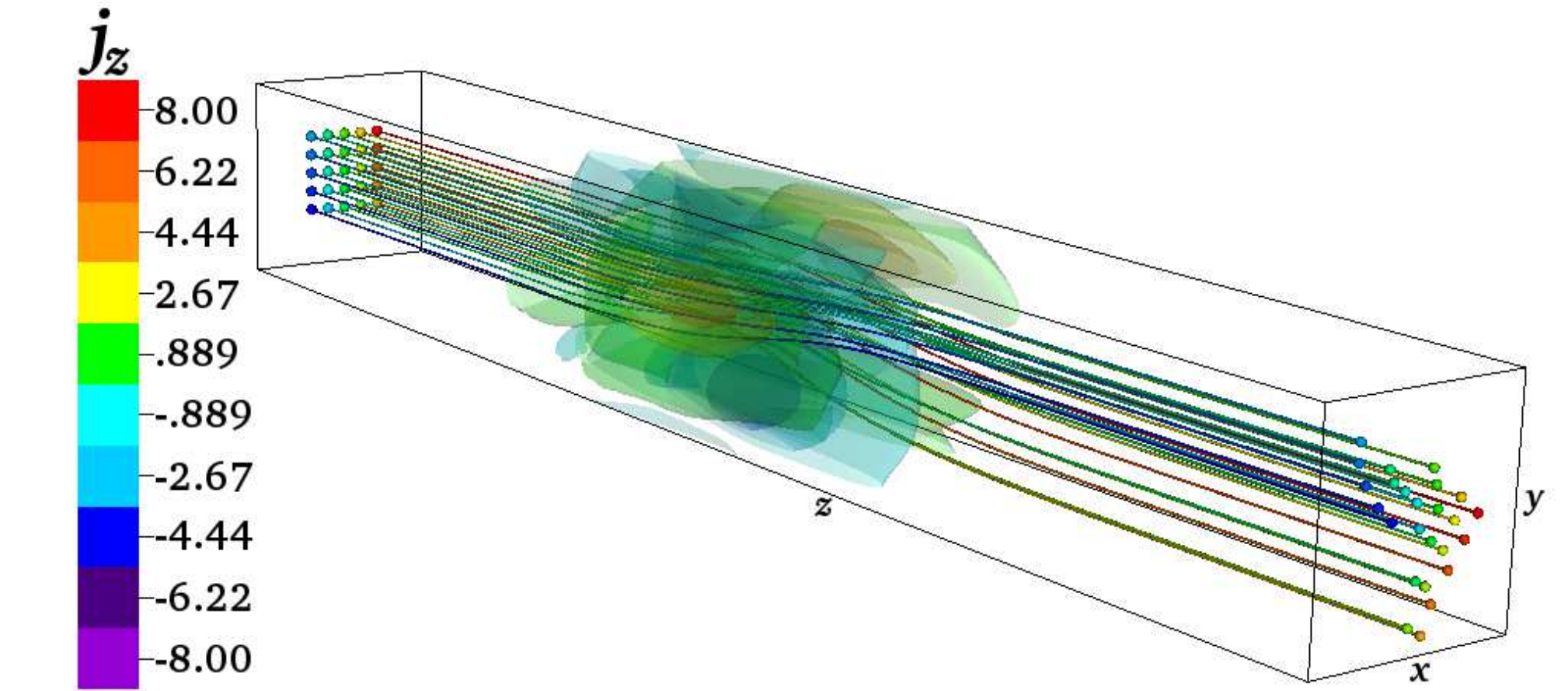}}                                   
\caption{Three-dimensional isocontours of the normalized parallel
  current density $j_z/j_0$ between \Alfven wavepacket collisions at
  (a) $t/T_c=0$, (c)  $t/T_c=1$, and  (e)  $t/T_c=2$ and at the midpoint of collisions at
  (b)  $t/T_c=0.5$, (d)  $t/T_c=1.5$, and  (f)  $t/T_c=2.5$.
\label{fig:3d}}
\end{figure}

\begin{figure}
\centering \includegraphics[scale = .7]{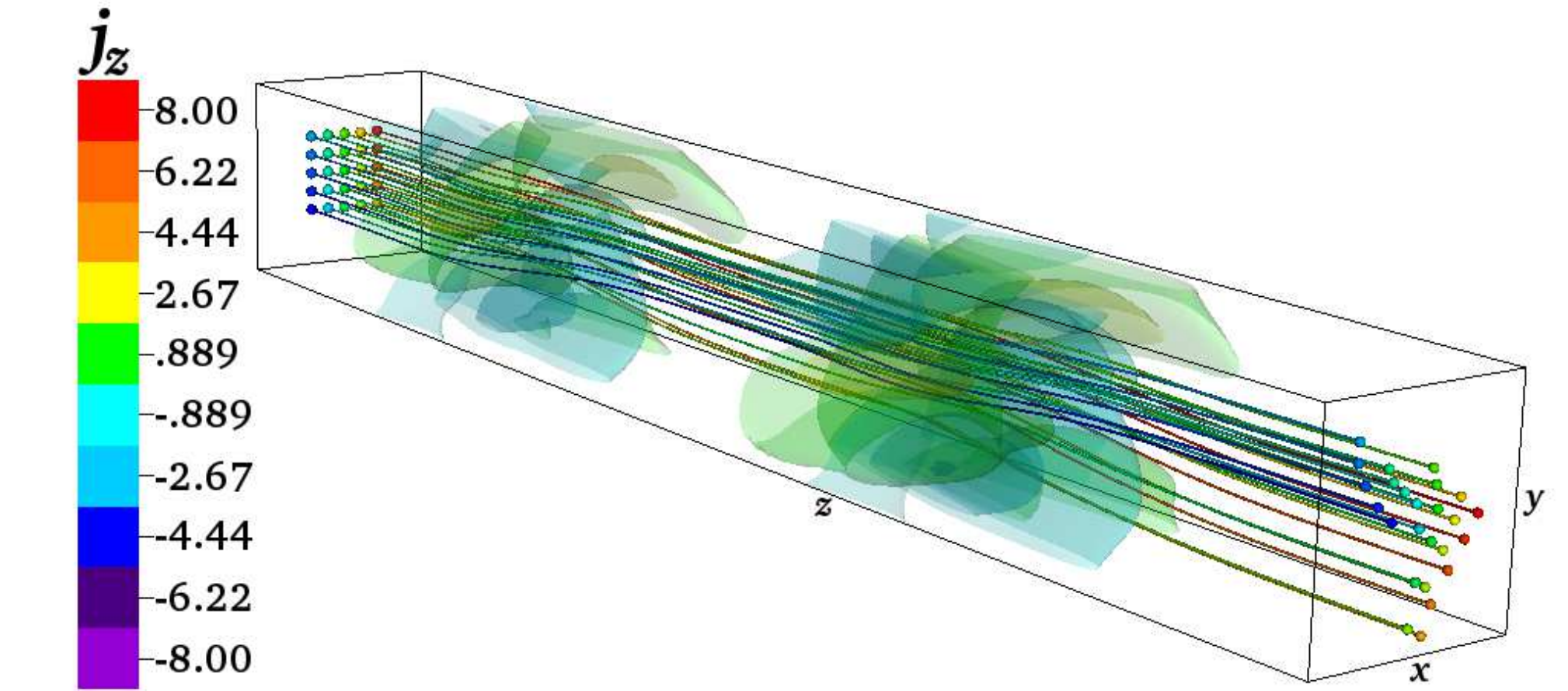}

\caption{3D movie of perpendicularly polarized
counterpropagating localized \Alfven wavepacket collisions. The rainbow
lines that extend along the length of the box represent magnetic field
lines and the rainbow contours represent isocontours of the current in
the $z$-direction, $J_{z}$. [Edited from original JPP submission, since arXiv does not support embedded media. Copy and paste the following link in your browser for access to the movie: https://www.dropbox.com/s/j161pt9j2t8r3wt/Movie1.mp4?dl=0 ]
\label{fig:mov}}
\end{figure}

The basic evolution of this strong \Alfven wavepacket collision
simulation is illustrated by three-dimensional contour plots of the
parallel current density $j_z$ associated with each of the interacting
wavepackets, shown in \figref{fig:3d} and displayed as a movie in \figref{fig:mov}. Time is normalized in terms of
the time for a single \Alfven wavepacket collision, $T_c$, during
which the initially separated \Alfven wavepackets approach each other
along $z$ (with the $+z$ direction from left to right in
\figref{fig:3d}), overlap and interact nonlinearly, and then move away
from each other after the collision.  At $t=0$, the unipolar
wavepacket is centered at $z=-L_z/4$, and the dipolar wavepacket at
$z=+L_z/4$; during a collision time $T_c$, each wavepacket propagates
at the \Alfven velocity $v_A$ over a parallel distance $L_z/2$.  The
midpoint of each collision occurs at $t/T_c=0.5, 1.5, 2.5$. 

In \figref{fig:3d}(a) are plotted isocontours of the normalized
parallel current density $j_z/j_0$ for each of the wavepackets at
$t=0$, where the current density is normalized by $j_0 = n_0q_i v_{ti}
L_\perp/L_z$. It is clear that the unipolar wavepacket (at $z<0$, left
side) initially has only perpendicular variation in the $x$ direction,
while the dipolar wavepacket (at $z>0$, right side) initially has only
perpendicular variation in the $y$ direction. \figref{fig:3d}(b) shows
the midpoint of the first collision occurring at $z=0$ and $t/T_c=0.5$,
showing a significantly more complicated perpendicular structure
parallel current density $j_z$.

After the first collision at $t/T_c=1.0$ in \figref{fig:3d}(c), the
unipolar wavepacket, now at $z=+L_z/4$, has gained some variation in
the $y$ direction, and the dipolar wavepacket, now at $z=-L_z/4$, has
developed variation in the $x$ direction. Each wavepacket has been
distorted by passing through, and interacting nonlinearly with, the
counterpropagating wavepacket. Mathematically, when expressed in terms
of Fourier modes, the strong \Alfven wavepacket collision has mediated
the nonlinear transfer of energy from the two initial perpendicular
Fourier modes to other Fourier modes with larger values of $k_\perp$,
as shown quantitatively in \secref{sec:perp}.  Therefore this
visualization clearly shows the nonlinear cascade of energy to smaller
scales in strong \Alfven wavepacket collisions, the fundamental
building block of astrophysical plasma turbulence, a key result of
this study.

Because the simulation domain is periodic in the $z$ direction, the
\Alfven wavepackets undergo a second collision at the boundary of the
domain $z=\pm L_z/2$ at $t/T_c=1.5$, shown in \figref{fig:3d}(d),
followed by a third collision at $z=0$ at $t/T_c=2.5$, shown in
\figref{fig:3d}(f). Below we explore in more detail the cascade of
energy to smaller perpendicular scales, the development of current
sheets, the evolution of the energy in perpendicular Fourier modes,
and key properties of localized \Alfven wavepacket collisions.

\subsection{The Perpendicular Cascade of Energy}
\label{sec:perp}
\begin{figure}

\centering {\includegraphics[trim=0 70bp 0 70bp,clip,scale=.7]{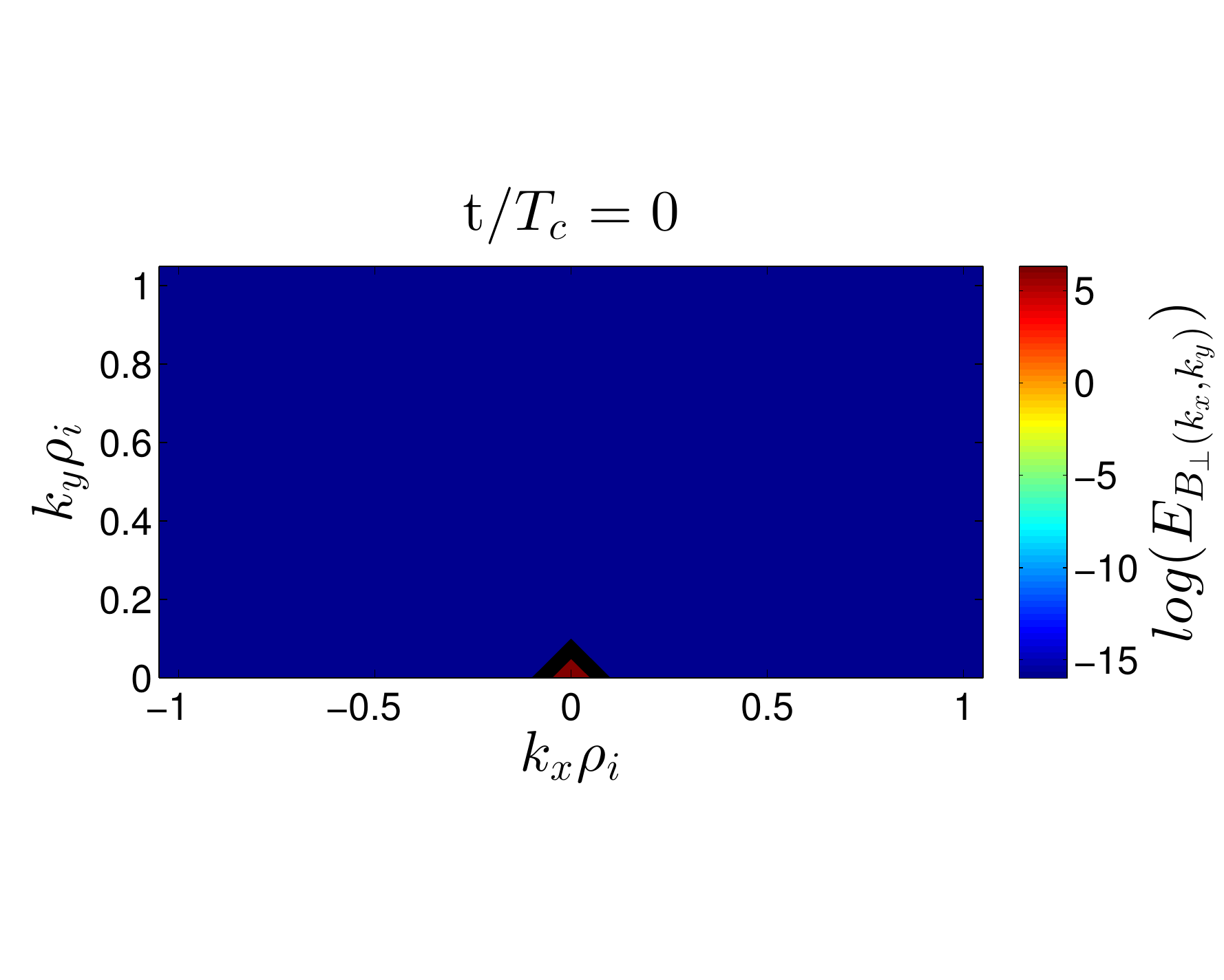}}

\centering {\includegraphics[trim=0 70bp 0 70bp,clip,scale=.7]{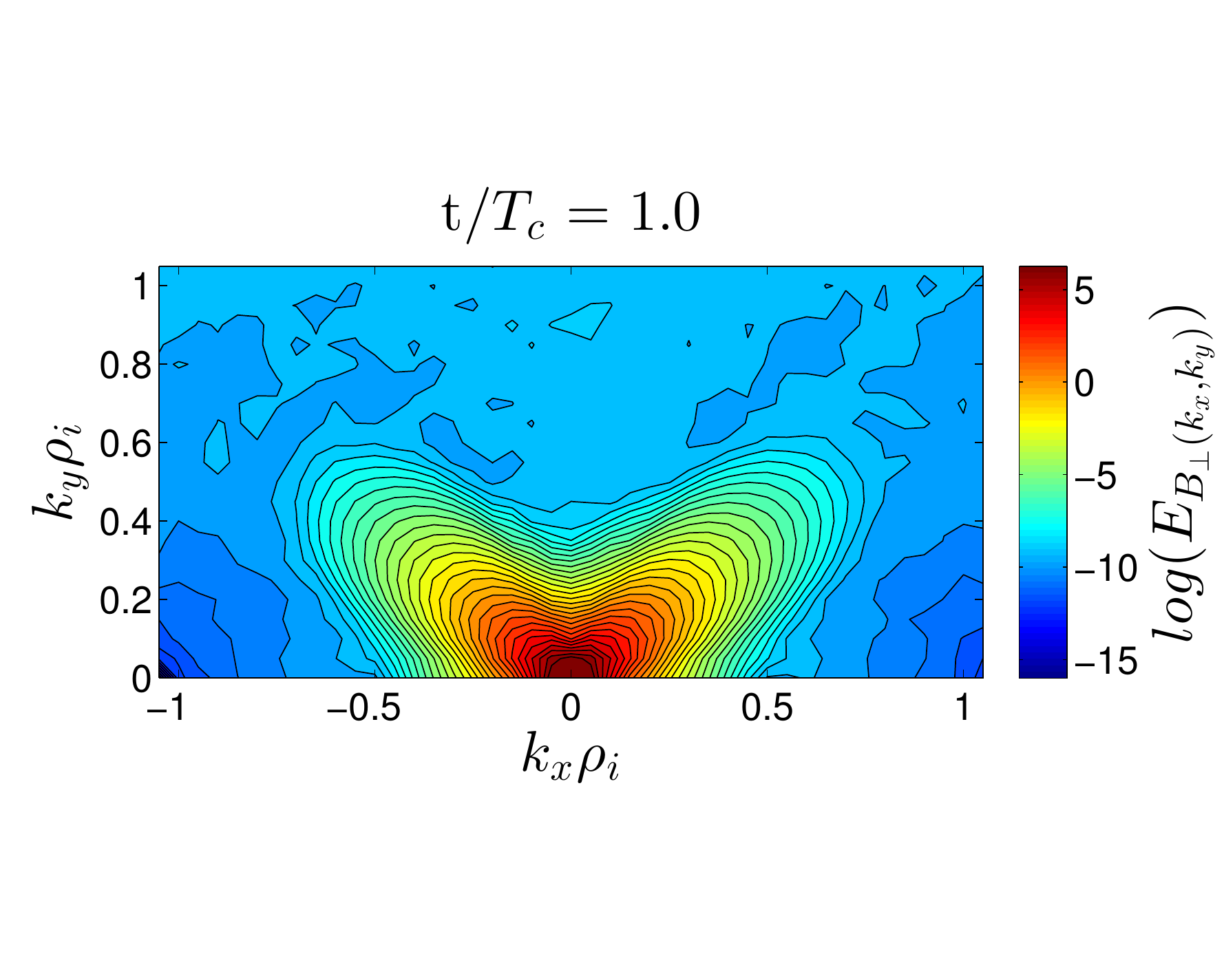}}

\centering {\includegraphics[trim=0 70bp 0 70bp,clip,scale=.7]{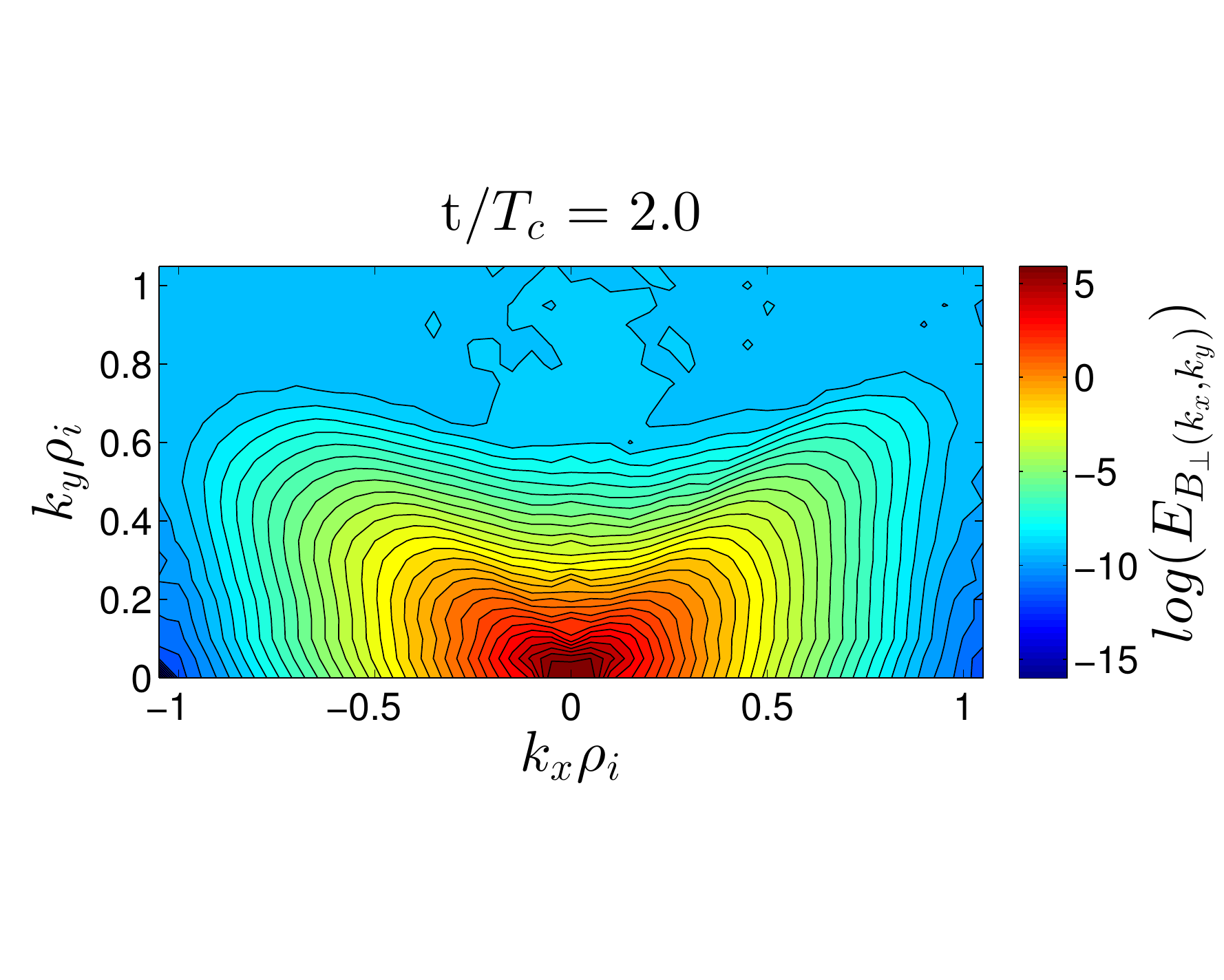}}

\caption{Plots of the perpendicular magnetic energy
 $E_{B_\perp(k_x,k_y)}$ (arbitrary units) on a log scale in the perpendicular
  Fourier plane $(k_x,k_y )$ (a) at the initial time $t/T_c=0$, (b) after the first
  strong \Alfven wavepacket collision at $t/T_c=1$, and (c) after the second
  strong \Alfven wavepacket collision at $t/T_c=2$.
\label{fig:power}}	
\end{figure}

To explore the nonlinear cascade of smaller perpendicular scales in
this strong \Alfven wavepacket collision simulation, we plot in
\figref{fig:power} the perpendicular magnetic energy of fluctuations
in perpendicular Fourier space integrated over $z$,
$E_{B_\perp(k_x,k_y)} \equiv \int_{-L_z/2}^{L_z/2} dz |\delta
B_\perp(k_x,k_y)|^2/8 \pi$ in arbitrary units. In
\figref{fig:power}(a) is the perpendicular magnetic energy of the
initial \Alfven wavepackets at $t=0$, showing all of the energy is
contained within the three perpendicular Fourier modes $(k_x
\rho_i,k_y \rho_i) = (1,0)$, $(0,1)$, and $(-1,0)$.  Note that
\T{AstroGK} uses a reality condition imposed on the complex Fourier
coefficients such that the magnetic field can be described using only
Fourier modes in the upper half-plane $k_y \ge 0$. The reality
condition requires $\delta \hat{B} (k_x,k_y) = \delta \hat{B}^*
(-k_x,-k_y)$, so if there is power in the $(1,0)$ mode, there must be
equivalent power in the $(-1,0)$ mode, as seen in \figref{fig:power}(a).

After the first collision at $t/T_c=1$, \figref{fig:power}(b) shows
definitively that the strong nonlinear interactions have transferred
significant energy to many other modes in the perpendicular Fourier
plane $(k_x,k_y )$.  Note that the perpendicular wavenumber is given
by $k_\perp = \sqrt{k_{x} + k_{y}}$, so that modes further away from
the origin represent smaller scale fluctuations in the perpendicular
plane. The plot after the second collision at $t/T_c=2$ in
\figref{fig:power}(c) shows that successive collisions continue to
facilitate the perpendicular cascade of energy. These results
demonstrate definitively that the finding from the idealized periodic
case---that \Alfven wave collisions mediate the nonlinear transfer of
energy to smaller perpendicular scales---indeed persists in the more
realistic case of localized \Alfven wavepacket collisions, answering
the first key question posed in the introduction. 
	
\subsection{Current Sheet Development}
\label{sec:cs}

\begin{figure}
\hspace{0.1in} (a) $t/T_c$=0.3 \hspace{1.5in}  (b) $t/T_c$=0.5
	{\includegraphics[scale=.4]{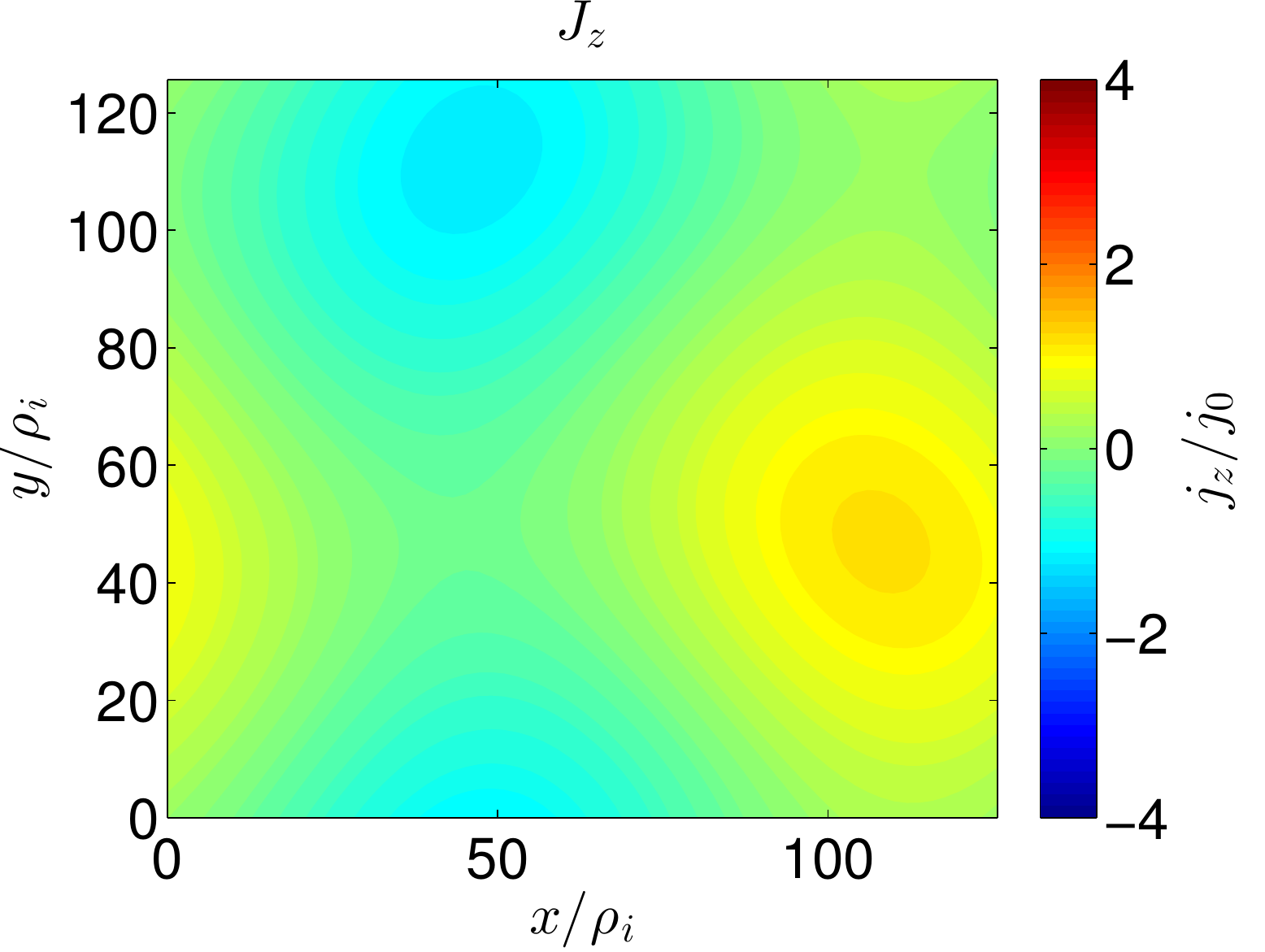}}\hfill 
	{\includegraphics[scale=.4]{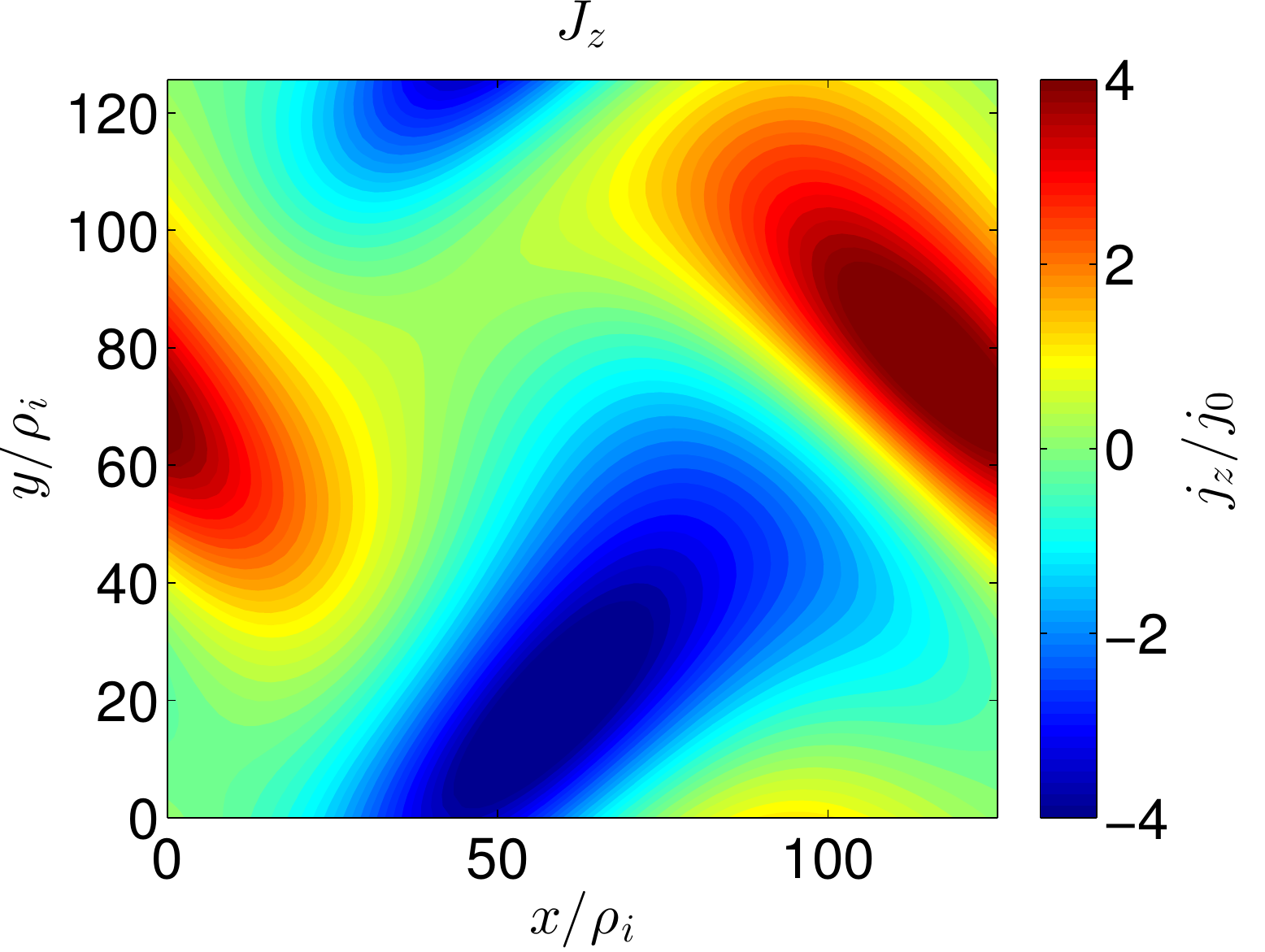}}
\vfill	
\hspace{0.1in} (c) $t/T_c$=1.3  \hspace{1.5in}  (d) $t/T_c $=1.5
	{\includegraphics[scale=.4]{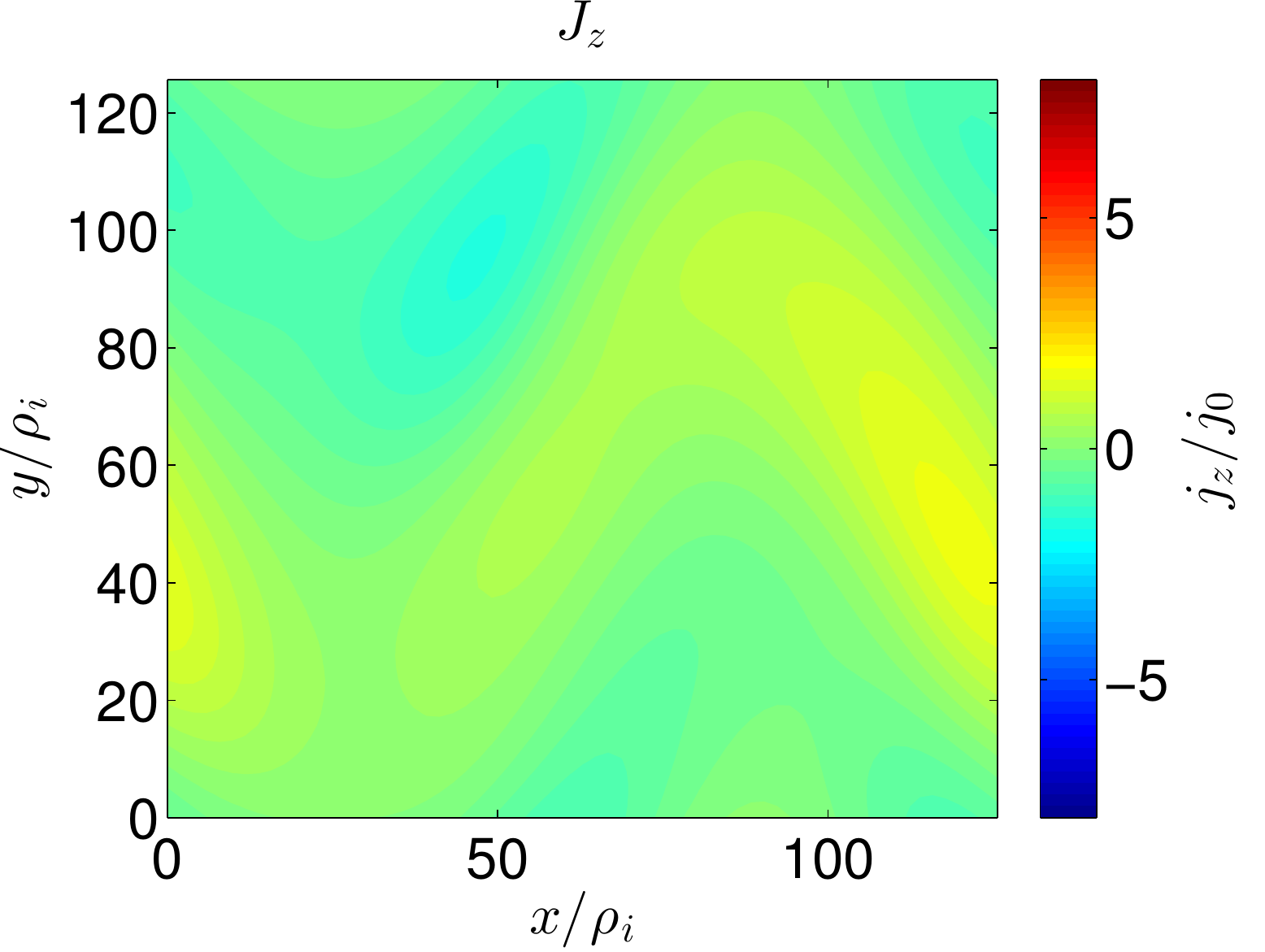}}\hfill
	{\includegraphics[scale=.4]{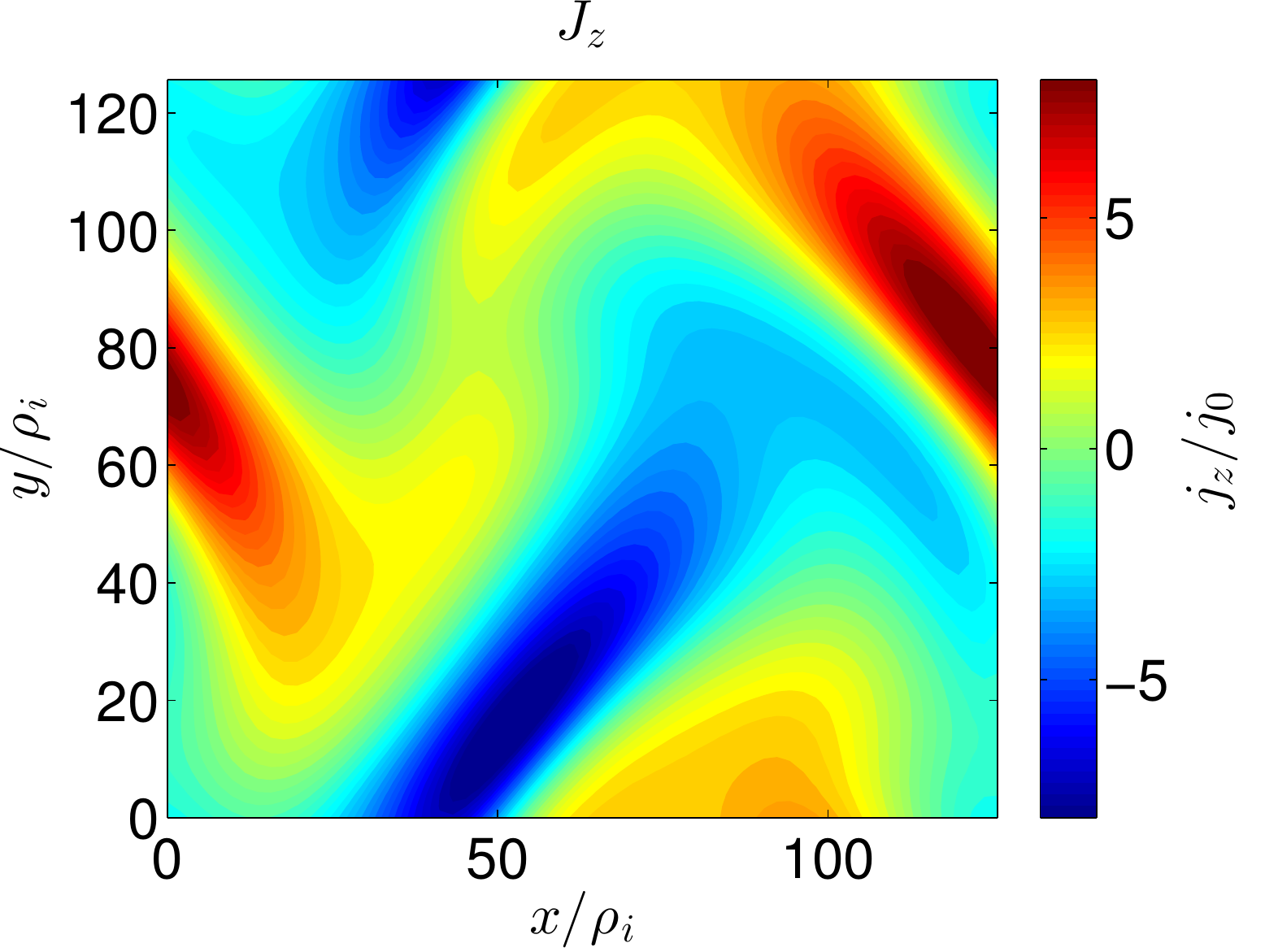}}
\vfill
\hspace{0.1in} (e) $t/T_c$=2.3  \hspace{1.5in}  (f) $t/T_c$=2.5
	{\includegraphics[scale=.4]{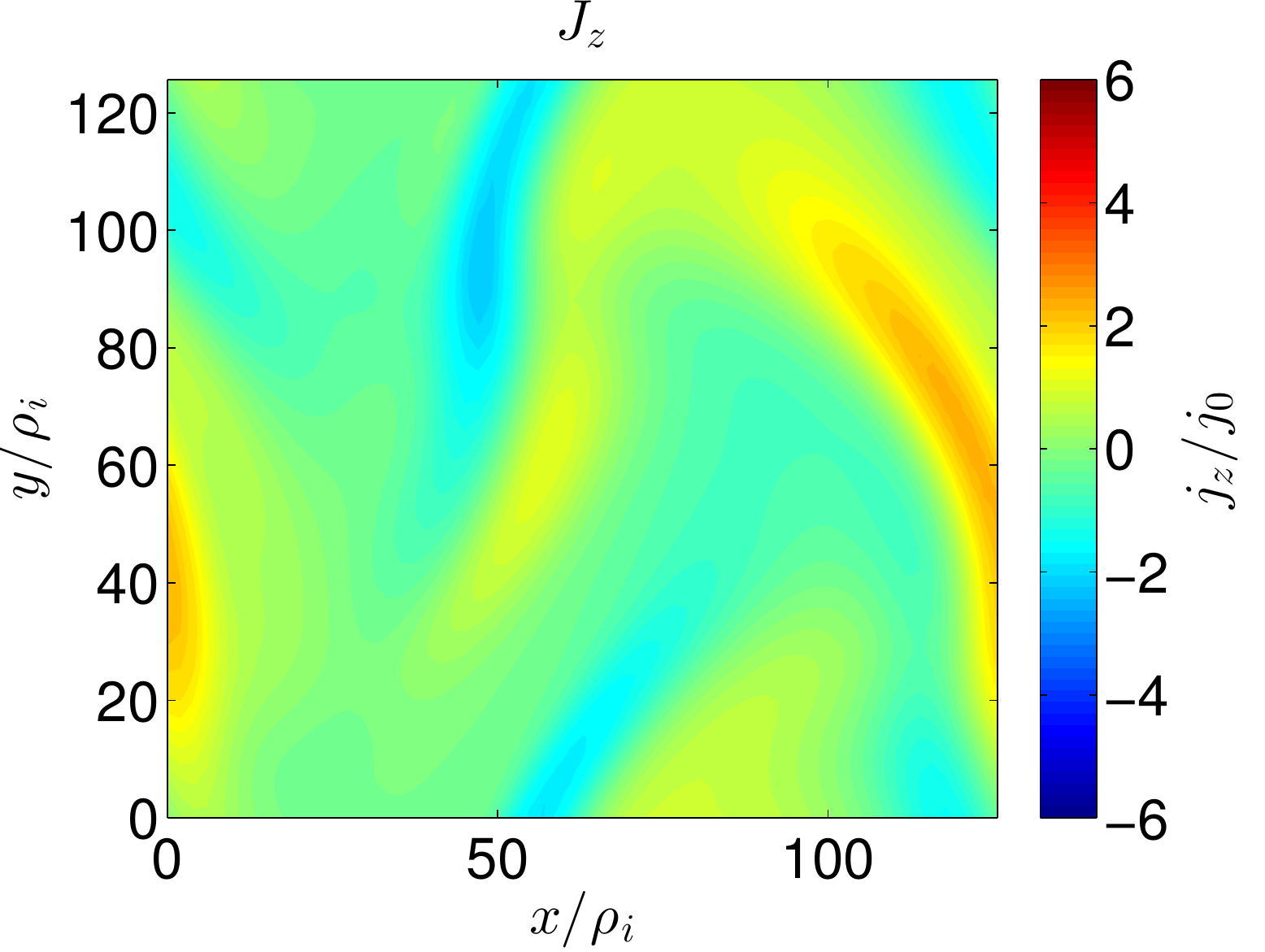}}\hfill
        {\includegraphics[scale=.4]{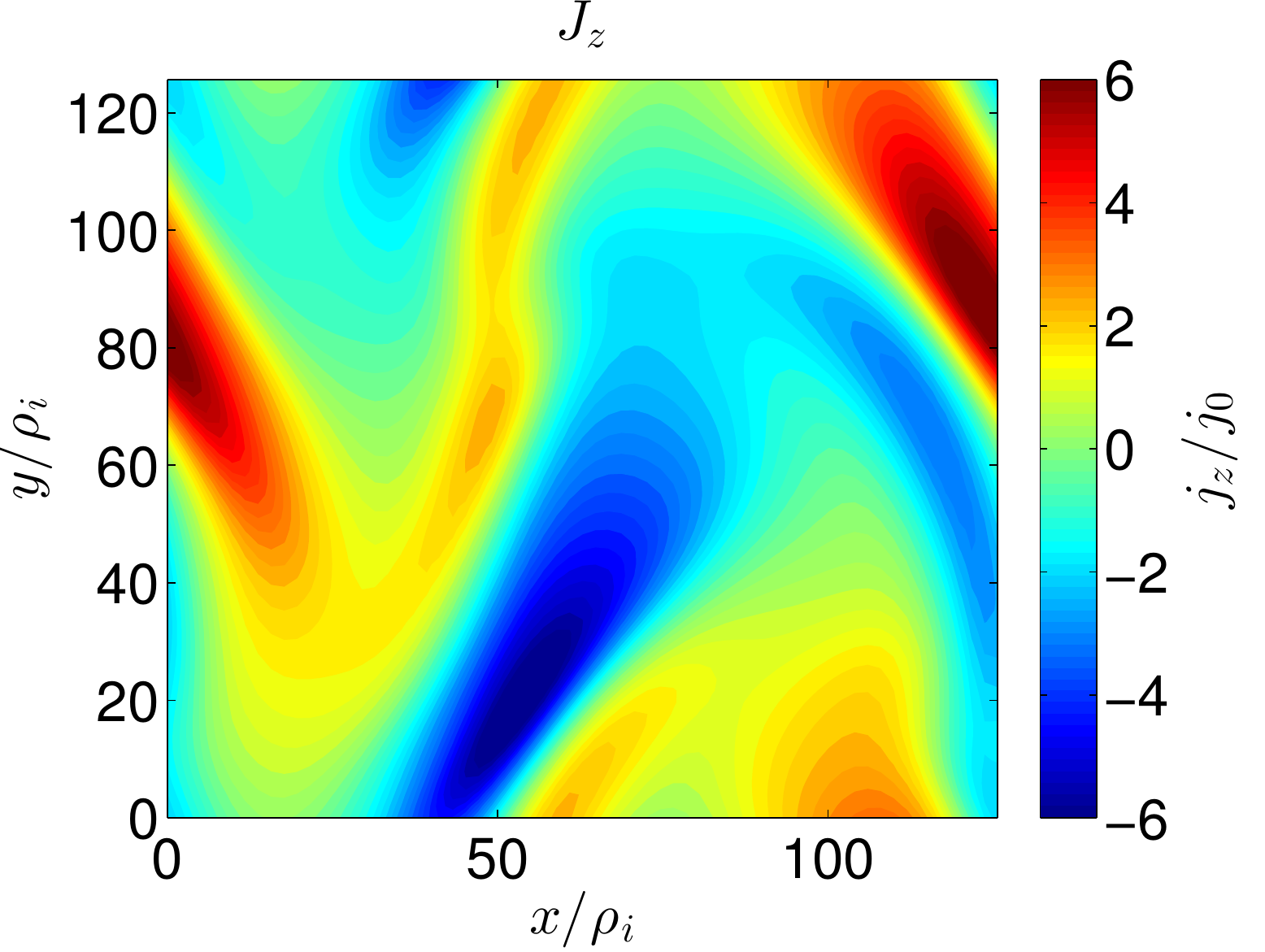}}	
\caption{Plot of the  normalized parallel current density
  $j_z/j_0$ at $z=0$  during the first collision at (a) $t/T_c=0.3$
  and (b)  $t/T_c=0.5$, at $z=L_z/2$ during second collision  at (c) $t/T_c=1.3$
  and (d)  $t/T_c=1.5$, and  at $z=0$  during the third collision at (a) $t/T_c=2.3$
  and (b)  $t/T_c=2.5$.
\label{fig:zcol}}	
\end{figure}

For plane \Alfven wave collisions in the idealized periodic case,
plots of the parallel current density $j_z$ in the $(x,y)$ plane
perpendicular to the equilibrium magnetic field demonstrate
that strong \Alfven wave collisions self-consistently generate current
sheets \citep{Howes:2016b}. These current sheets extended the full
parallel length of the original \Alfven waves, with widths in the
perpendicular plane of approximately the perpendicular wavelength of
the original interacting \Alfven waves, but with a much smaller thickness in
the perpendicular plane. Here we use plots of the $(x,y)$ plane
perpendicular to the equilibrium magnetic field to determine whether
this current sheet formation persists in the more realistic case of
strong collisions between localized \Alfven wavepackets.

\begin{figure}
\hspace{0.1in} (a) $t/T_c$=0.0 \hspace{1.5in}  (b) $t/T_c$=0.0
	{\includegraphics[scale=.4]{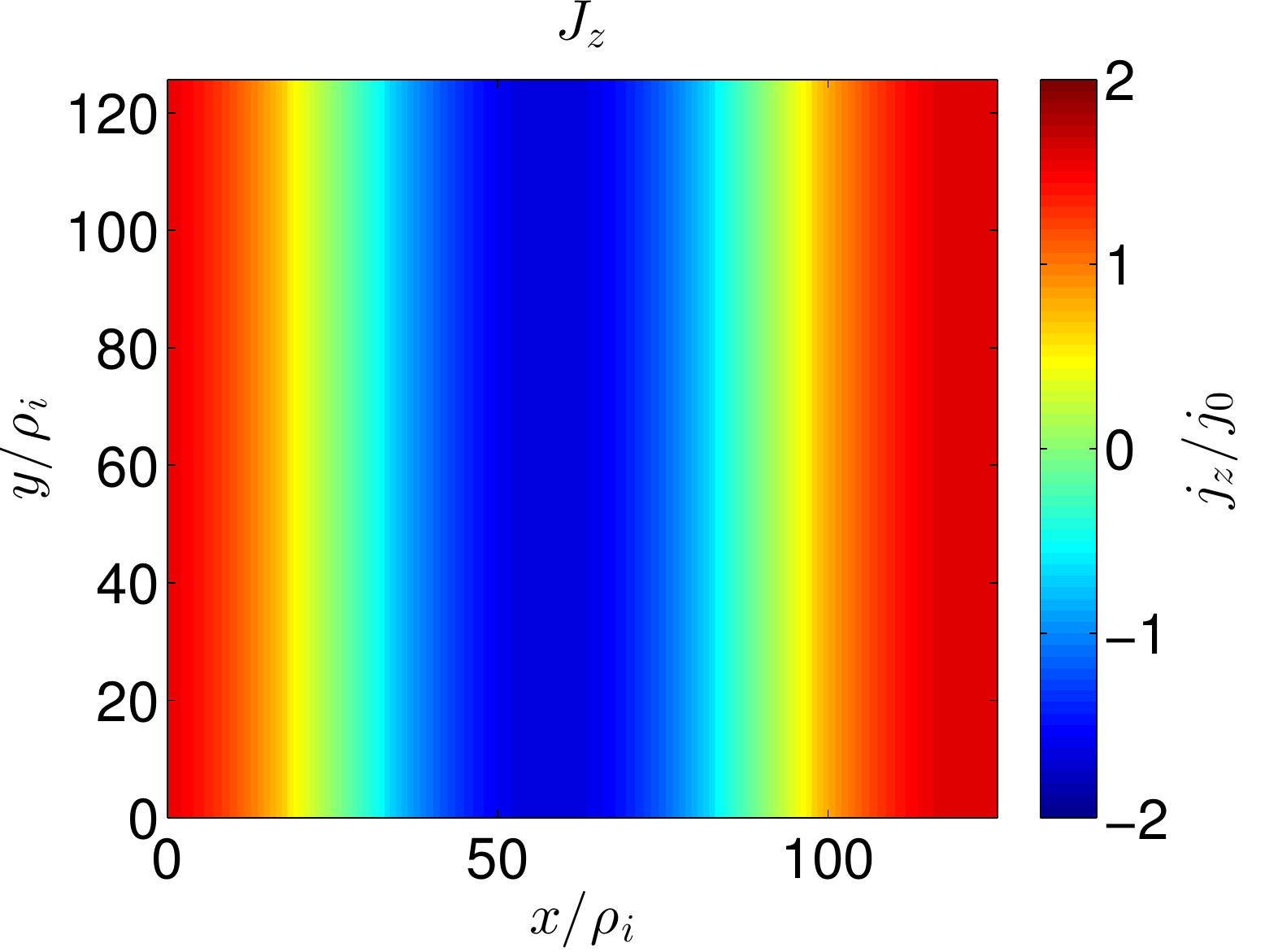}}\hfill
	{\includegraphics[scale=.4]{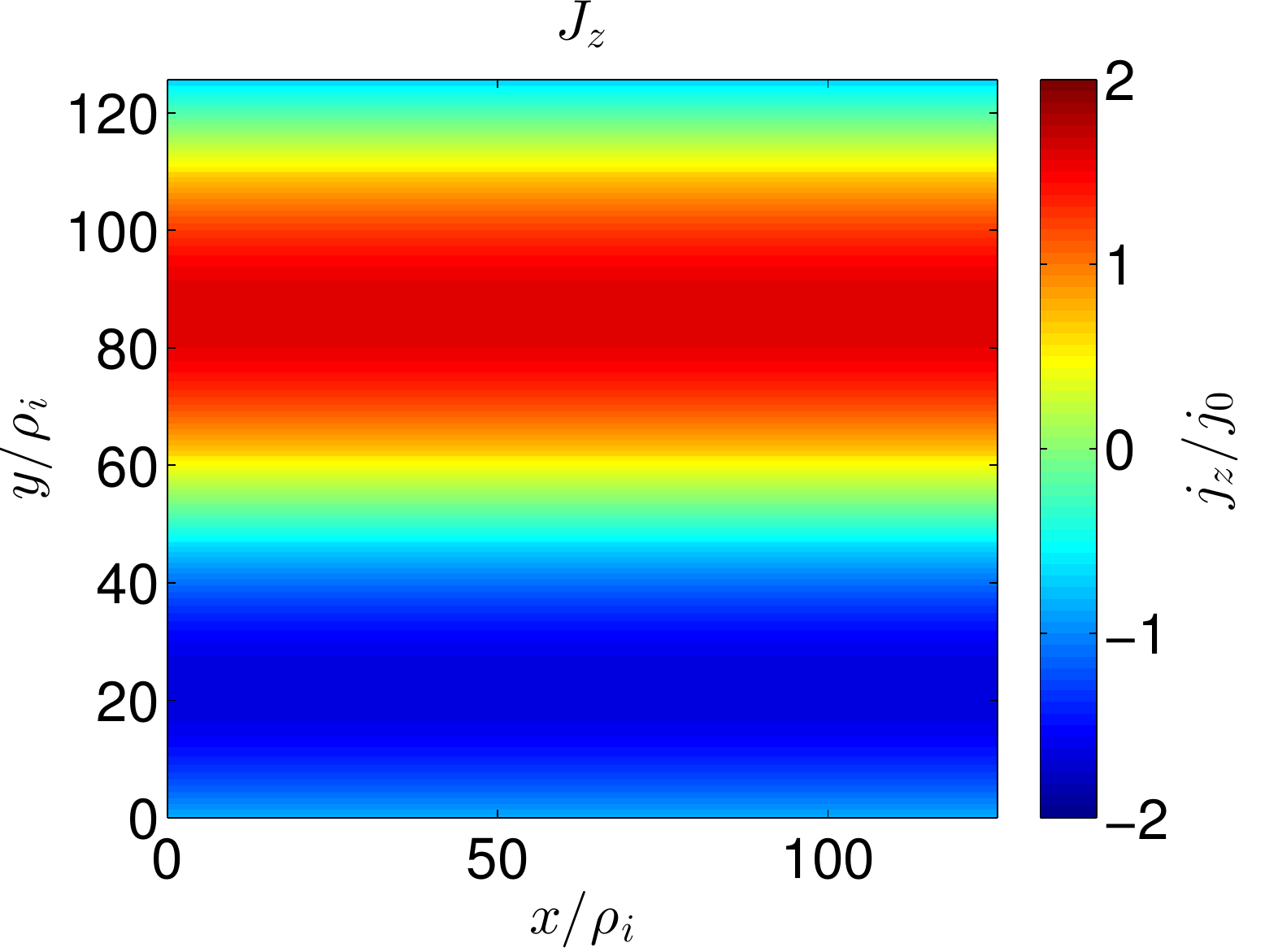}}
\vfill
\hspace{0.1in} (c) $t/T_c$=0.98 \hspace{1.5in}  (d) $t/T_c$=0.96
	{\includegraphics[scale=.4]{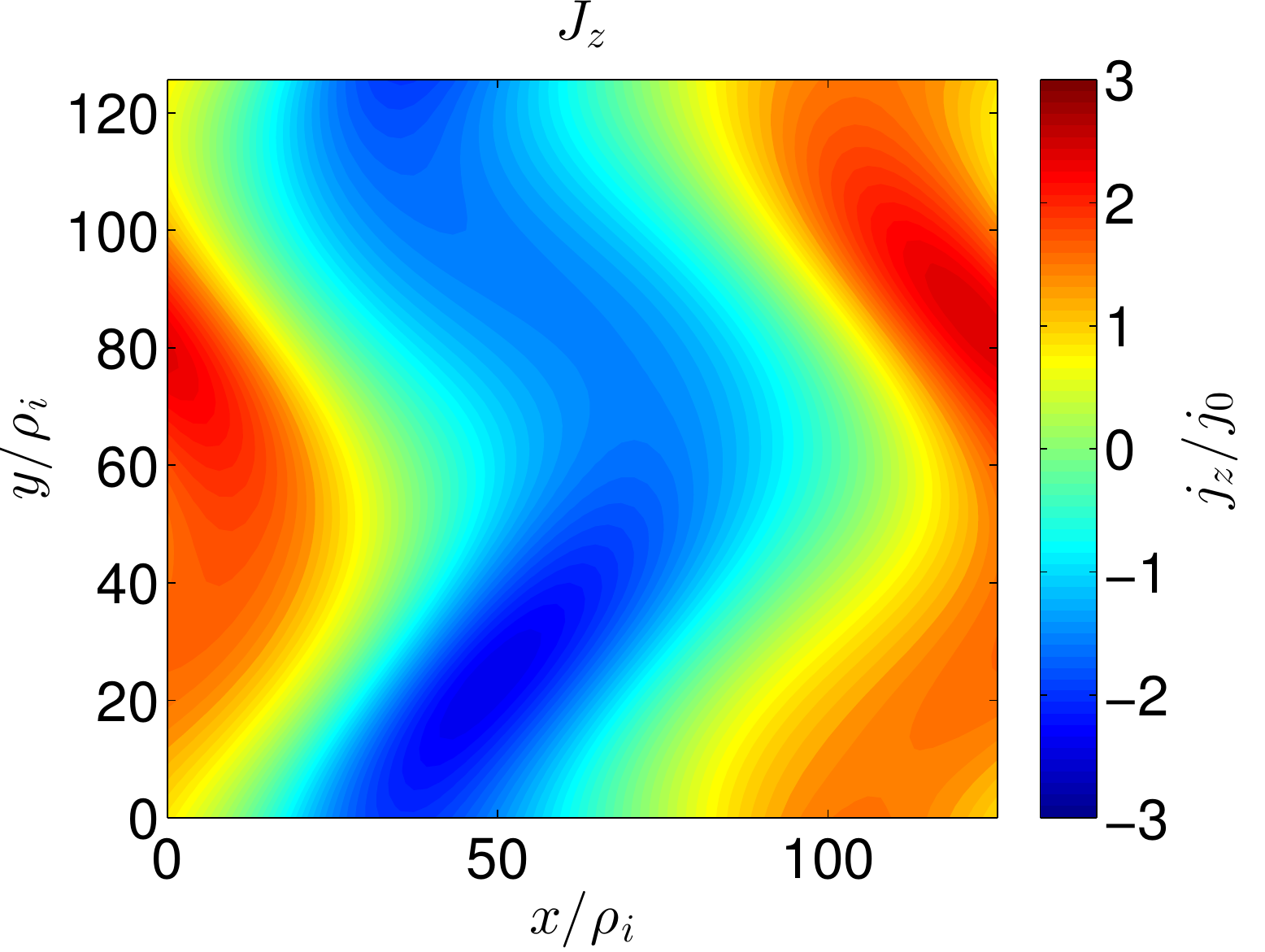}}\hfill
	{\includegraphics[scale=.4]{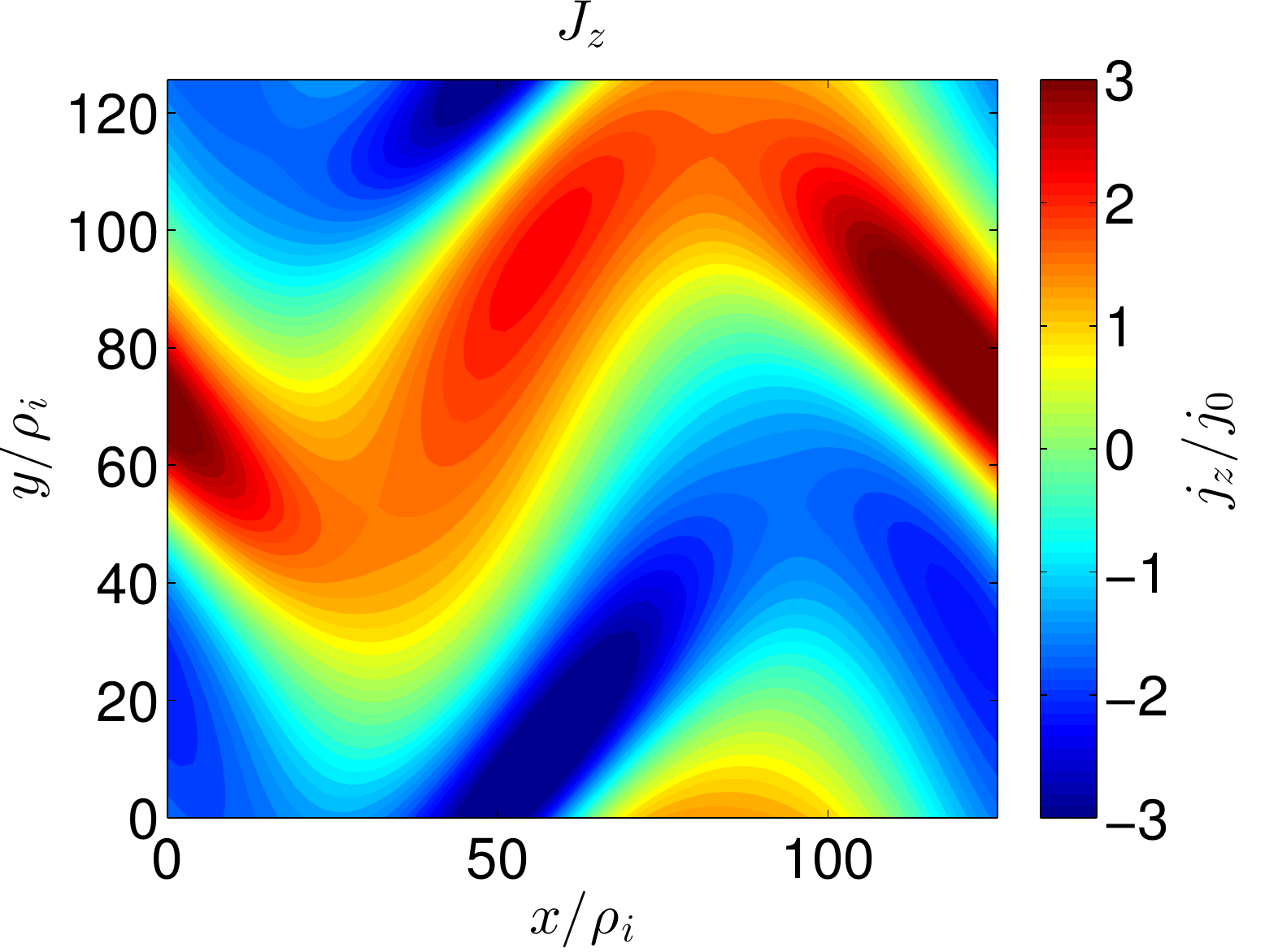}}
\vfill
\hspace{0.1in} (e) $t/T_c$=2.1 \hspace{1.5in}  (f) $t/T_c$=2.0	
        {\includegraphics[scale=.4]{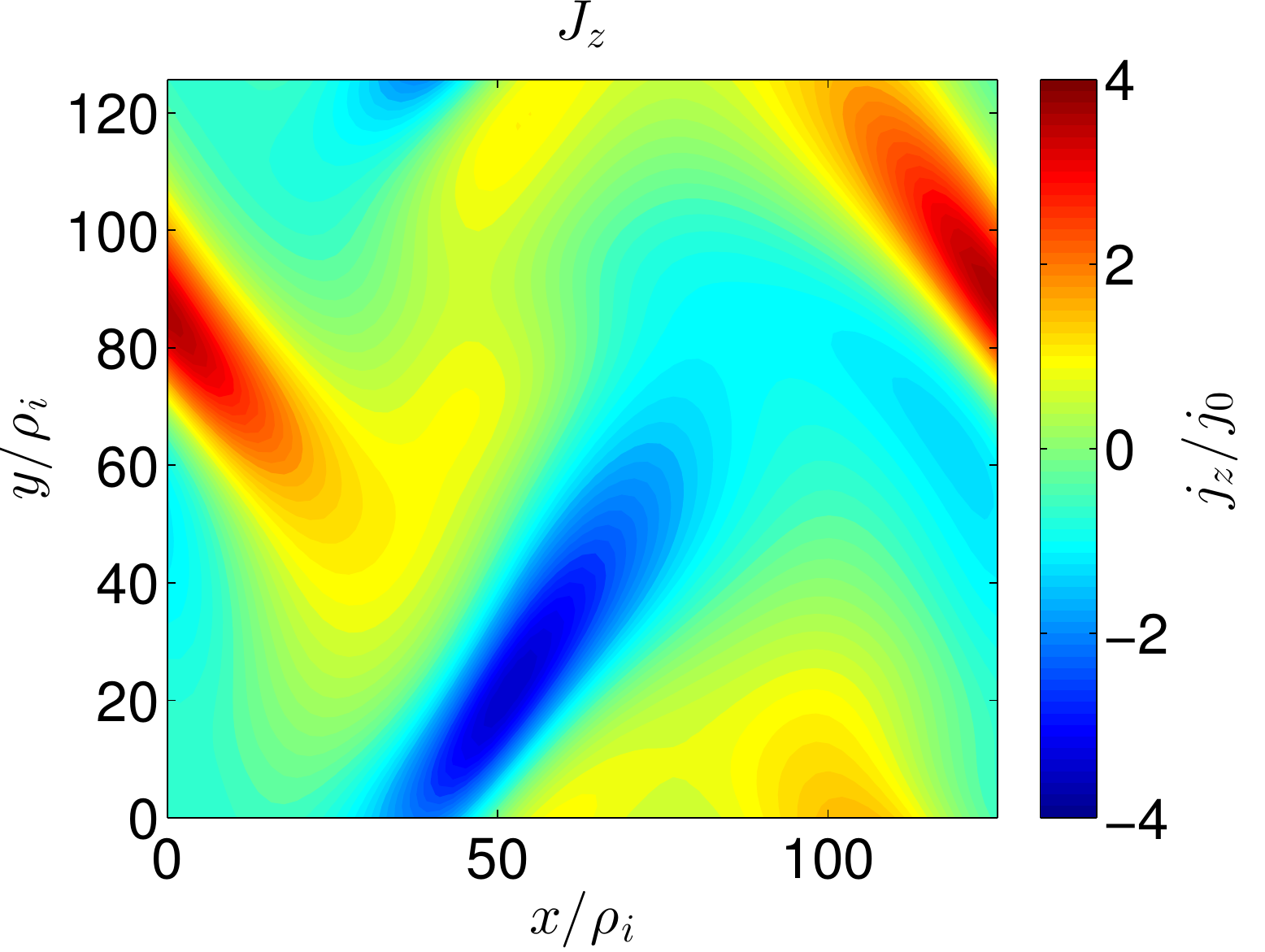}}\hfill	
        {\includegraphics[scale=.4]{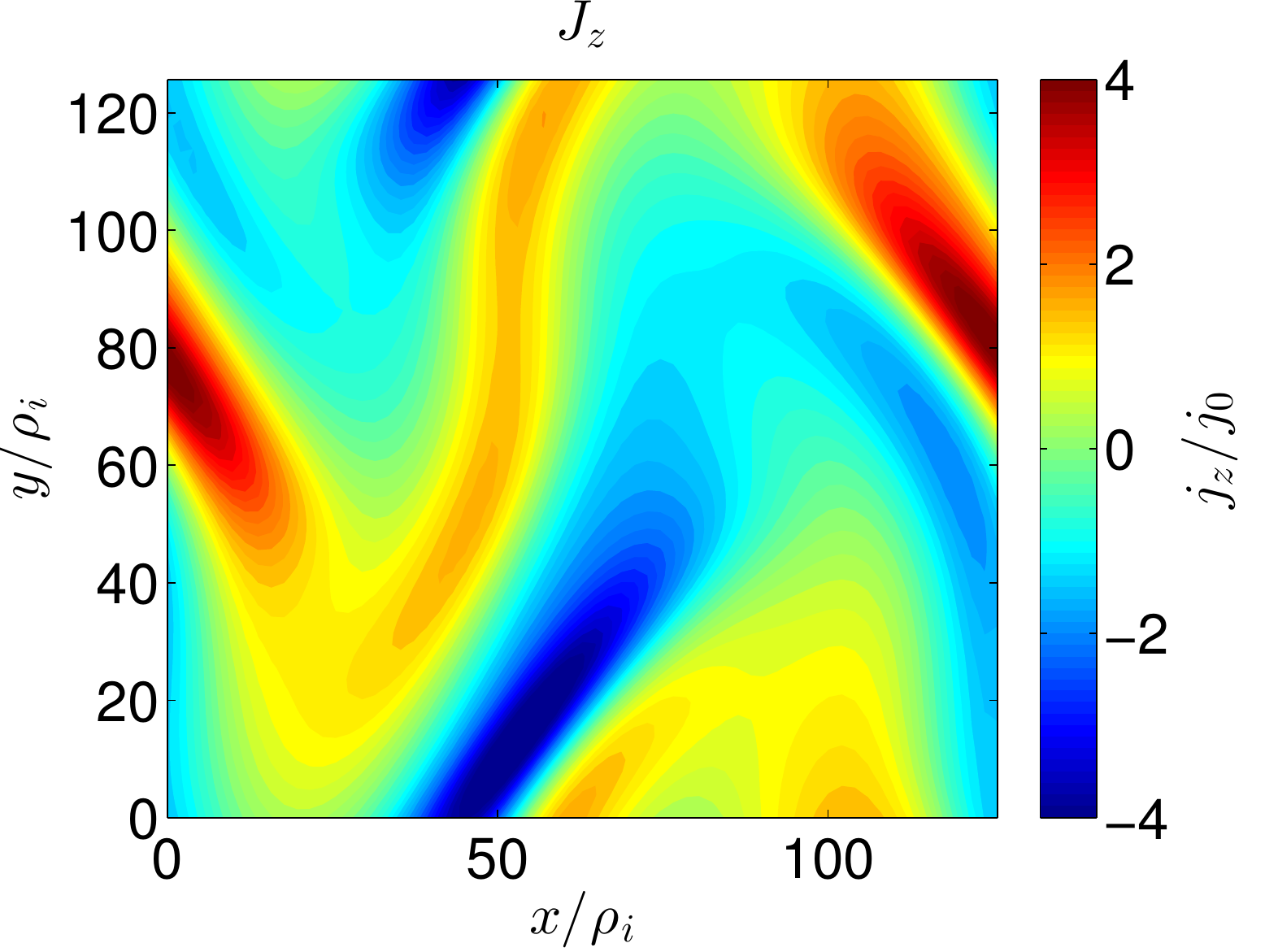}} 
\caption{Plot of the normalized parallel current density $j_z/j_0$ of
  colliding \Alfven wavepackets before the first collision at $t=0$
  for (a) the unipolar wave at $z=-L_z/4$ and and (b) the dipolar wave
  at $z=+L_z/4$, after the first collision when the wavepackets have separated for (c) 
   the unipolar wave at $z=+L_z/4$ and $t/T_c=0.98$  and (d)
   the dipolar wave at $z=-L_z/4$ and $t/T_c=0.96$, and after the second collision
   for (e) the unipolar wave
at $z=-L_z/4$ and $t/T_c=2.1$ and (f) the dipolar wave at $z=+L_z/4$
and $t/T_c=2.0$.
\label{fig:zba}}	
\end{figure}

First, we examine the development of the parallel current density
$j_z$ in the $(x,y)$ plane during the course of each collision.  In
\figref{fig:zcol}, we plot the normalized parallel current density
$j_z/j_0$ at $z=0$ (a) early in the first collision at $t/T_c=0.3$
and (b) later in the same collision at $t/T_c=0.5$, where we remind
the reader that the midpoint of each collision occurs at $t/T_c=0.5, 1.5, 2.5$.
The nearly circular $j_z$ pattern in \figref{fig:zcol}(a) is largely
due to the linear superposition of the current of each of the initial
\Alfven wavepackets (the top row of \figref{fig:zba} shows the current
patterns of each of the initial \Alfven wavepackets). The nonlinear
interaction of the \Alfven wavepackets during the first collision
leads to the thinning of these initial circular current patterns into
a more sheet-like morphology in \figref{fig:zcol}(b). The second row
of \figref{fig:zcol} shows the evolution of the second collision in
the plane $z=L_z/2$ (at the parallel boundary of the periodic domain)
at (c) $t/T_c=1.3$ and (d) $t/T_c=1.5$.  Here we see the current
sheets becoming yet more elongated and intense through the course of
the second collision. The third row shows plots of $j_z/j_0$ during
the third collision at $z=0$, with further thinning and
intensification of the current sheets. Note that the three-dimensional
isocontours of $j_z$ in \figref{fig:3d} shows that the parallel extent
of the current sheets is the same as the parallel length of the
original wavepackets.

These results show clearly that strong collisions of localized \Alfven
wavepackets indeed self-consistently generate intermittent current
sheets, definitively answering the second question in the
introduction.

Another important question, however, is whether the current sheets
that develop during the wavepacket collisions persist within each
wavepacket after the collision.  In \figref{fig:zba}, we show the
initial pattern at $t=0$ of the normalized parallel current density
$j_z/j_0$ for (a) the unipolar wave at $z=-L_z/4$ with only variation
in the $x$ direction and (b) the dipolar wave at $z=+L_z/4$ with only
variation in the $y$ direction.  After the first collision, we show
$j_z$ for (c) the unipolar wave at $z=+L_z/4$ and $t/T_c=0.98$ and (d)
the dipolar wave at $z=-L_z/4$ and $t/T_c=0.96$.  At this time when
the wavepackets are no longer overlapping in $z$, one can clearly see
the distortion of current density pattern, due to the previous
collision, has lead to a thinning and intensification of the current density into a more sheet-like morphology that persists after the wavepackets have separated.  After the second collision, we plot (e) the unipolar wave
at $z=-L_z/4$ and $t/T_c=2.1$ and (f) the dipolar wave at $z=+L_z/4$
and $t/T_c=2.0$.  Here we can see that the current sheets that develop
as a consequence of the strong \Alfven wavepacket collisions indeed
persist within the wavepackets after the collision is over, showing
that these interactions in the more realistic case may explain the
ubiquitous observations of intermittent current sheets in plasma
turbulence.

\subsection{Evolution of Energy}
\label{sec:energy}

A simple physical interpretation of the evolution of the energy in
this localized \Alfven wavepacket collision can be developed using the
equations of incompressible MHD for guidance. These equations, written
in terms of Elsasser variables, take the form
\begin{equation}
\frac{\partial \V{z}^{\pm}}{\partial t} 
\mp \V{v}_A \cdot \nabla \V{z}^{\pm} 
=-  \V{z}^{\mp}\cdot \nabla \V{z}^{\pm} -\nabla P/(n_{0i}m_i+n_{0e}m_e),
\label{eq:elsasserpm}
\end{equation}
and $\nabla\cdot \V{z}^{\pm}=0$.  Here $\V{v}_A =\V{B}_0/\sqrt{4
  \pi(n_{0i}m_i+n_{0e}m_e)}$ is the \Alfven velocity due to the
equilibrium field $\V{B}_0=B_0 \zhat$ where $\V{B}=\V{B}_0+ \delta
\V{B} $, $P$ is total pressure (thermal plus magnetic), and
$n_{0i}m_i+n_{0e}m_e$ is mass density.  Recall that the Elsasser
variables are defined by $\V{z}^{\pm} = \V{u} \pm \delta
\V{B}/\sqrt{4 \pi (n_{0i}m_i+n_{0e}m_e)}$, representing waves that
propagate up or down the mean magnetic field. The nonlinear term,
$\V{z}^{\mp}\cdot \nabla \V{z}^{\pm} $, governs the nonlinear
interactions of the counterpropagating \Alfven wave collisions.

\begin{figure}
\centering \includegraphics[scale=.5]{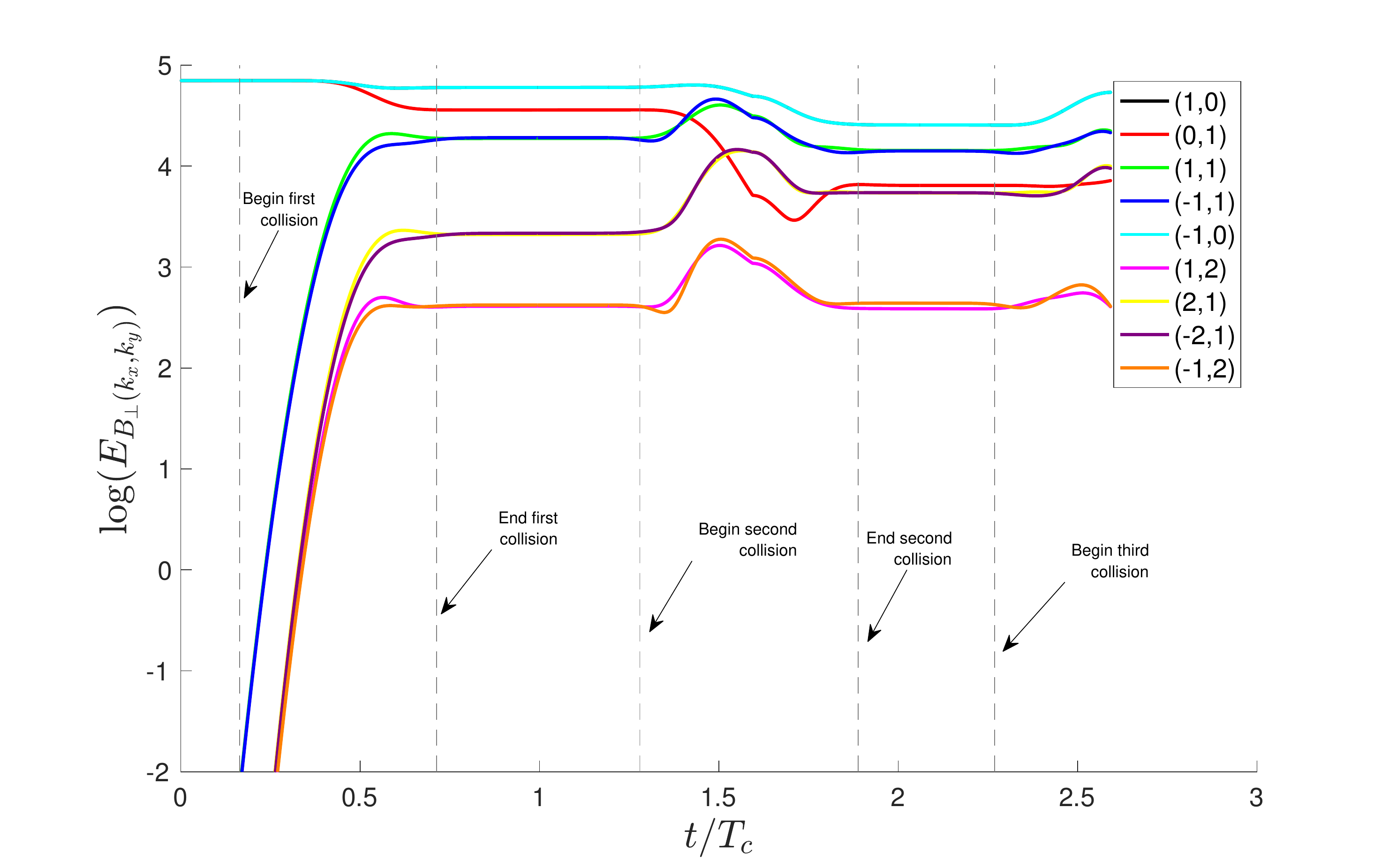}
\caption{Plot of the evolution of energy transfer between
  $(k_{x},k_{y})$ modes. Note that the (1,0) mode is plotted as the
  same line as the cyan (-1,0) mode, indicating they are identical.
 \label{fig:energy}}
 \end{figure}

In \figref{fig:energy}, we plot the temporal evolution of the magnetic
energy $E_{B_\perp(k_x,k_y)}$ in some of the lowest perpendicular
Fourier modes $(k_{x},k_{y})$, illustrating some of the key properties
of \Alfven wavepacket collisions.  The first main point is that there
is no nonlinear transfer of energy among Fourier modes until the two
wavepackets begin to overlap in space along $z$ (first vertical dashed
line).  As expected from theoretical considerations of incompressible
MHD turbulence
\citep{Iroshnikov:1963,Kraichnan:1965,Sridhar:1994,Montgomery:1995,Ng:1996,Galtier:2000,Howes:2013a},
the nonlinear interaction is zero unless both $z^-$ (the unipolar
wavepacket) and $z^+$ (the dipolar wavepacket) are both non-zero at
the same point in space, as can be seen by inspection of the nonlinear
term $\V{z}^{\mp}\cdot \nabla \V{z}^{\pm} $. Between collisions, when
the wavepackets have no overlap in $z$, there is no nonlinear transfer
of energy among different $(k_{x},k_{y})$ modes, as clearly seen in
\figref{fig:energy}.  The second main point is that the perpendicular
Fourier modes $(k_x L_\perp,k_y L_\perp)$ of the initial \Alfven
wavepackets, $(1,0)$ and $(0,1)$ generally lose energy to nonlinearly
generated perpendicular Fourier modes during each collision.  This is
the transfer of energy to smaller perpendicular scales, illustrated by
the plots of perpendicular magnetic energy in $(k_{x},k_{y})$ Fourier
space in \figref{fig:power}.  A more quantitative examination
contrasting the nonlinear transfer of energy between the idealized
plane \Alfven wave collisions in a periodic geometry and localized
\Alfven wavepackets collisions, in both the weakly and strongly
nonlinear limits, will be presented in a subsequent paper
\citep{Verniero:2017b}.

\subsection{Wavepacket Collisions}
\label{sec:properties}
The three-dimensional visualization of this localized \Alfven
wavepacket collision in \figref{fig:3d} presents a concise overview of
many of the properties examined above.  The series of panels (a)
through (c) show clearly the distortion of the original wavepackets
that arises during the collision and persists after the wavepackets
have separated again. This is a physical visualization of the nonlinear cascade of energy to smaller scales, the key characteristic 
of turbulence in space and astrophysical plasmas. Although a little
difficult to see clearly in the 3D projection in \figref{fig:3d}, the
nonlinear evolution of this strong \Alfven wavepacket collision also
leads to the self-consistent generation of a sheet-like morphology for
the regions of intense parallel current density $j_z$. This thinning
and intensifying of the current into sheets is more clearly seen in
the perpendicular cross section at $z=L_z/2$ (the right-hand $z$
boundary of the simulation domain) in \figref{fig:3d}(d); the same
cross section at $z=L_z/2$ is also shown in \figref{fig:zcol}(d).

In addition to the isocontours of the normalized parallel current
density $j_z/j_0$, we also plot the paths of a sample of magnetic
field lines as they traverse the domain.  Beginning at points on a $5
\times 5$ grid at $z=-L_z/2$, we trace the field lines in the $+z$
direction.  Although the perpendicular magnetic field fluctuation
associated with each \Alfven wavepacket is small compared to the
equilibrium magnetic field, $\delta B_\perp \ll B_0$, one can still
see the distortion of the field lines as they pass through each
wavepacket. In the space between wavepackets, both before and after
each collision, the magnetic field lines are straight and uniform,
further illustrating the point that nonlinear evolution ceases when
the wavepackets are separated, even in a gyrokinetic simulation where
dispersive and kinetic effects are resolved.

It is worthwhile also noting the important point that, even after each
strong \Alfven wavepacket collision, the individual wavepackets
continue to propagate along the equilibrium magnetic field and they
remain localized in $z$.  There is a very small spreading of the
wavepacket, hardly noticeable in \figref{fig:3d}, due to the fact that
\Alfven waves become dispersive at $k_\perp \rho_i \gtrsim 1$, with an
increasing parallel group velocity.  Therefore, some of the higher
$k_\perp$ modes that are nonlinearly generated by the strong \Alfven
wavepacket collision in this gyrokinetic simulation---a numerical
approach that resolves these dispersive kinetic effects at $k_\perp
\rho_i \gtrsim 1$---will propagate a little faster than \Alfven waves
in the MHD limit $k_\perp \rho_i \ll 1$ that have a parallel group
velocity at the \Alfven speed $v_{g \parallel} = v_A$, causing a
slight spreading out of the localized wavepacket.

\subsection{Relation to Similar Wavepacket Collision Simulations}
A recent study by \citet{Pezzi:2017} has used numerical simulations to
tackle what they term the ``Parker-Moffatt problem.'' The basic
question is inspired by the properties of incompressible MHD equations
given in \eqref{eq:elsasserpm}: do two initially separated, localized
\Alfven wavepackets interact nonlinearly only during the time that
they overlap, and then cease evolving after they have passed through
each other and become separated again?  The simulation results shown in
\figref{fig:3d}, and examined in further detail in the subsections
above, definitively answer this question---yes.

The results of the gyrokinetic simulation of strong \Alfven wavepacket
collisions presented here, however, stand in stark contrast to the
results of the study by \citet{Pezzi:2017}, which finds that after the
collision, in both their compressible MHD and Hybrid Vlasov-Maxwell
simulations, the wavepackets do not separate cleanly and continue to
evolve, as illustrated in Figure~1 of their paper. What is the cause
for the dramatic contradiction in these results?

We argue here that the limitation of the \citet{Pezzi:2017}
simulations to two spatial dimensions is the root of these striking
differences.  Due to the large computational cost of running the
Hybrid Vlasov-Maxwell (HVM) code \citep{Valentini:2007} in three
spatial dimensions, the authors chose to approximate the 3D problem by
using an oblique 2D approach \citep{Howes:2014c}, where spatial
variation is allowed in two dimensions in the $(x,y)$ plane, but the
equilibrium magnetic field is slightly tilted away from the normal to
the plane, $\V{B}_0 = B_0(\sin \theta \xhat + \cos \theta \zhat)$ with
$\theta = 6^\circ$, yielding a small component of the equilibrium
magnetic field in the $x$ direction. Since plane \Alfven waves only
propagate if there is a nonzero component of the wavevector parallel
to the equilibrium magnetic field, $k_\parallel \ne 0$, this trick
enables the $x$ component of a wavevector to have a small component
parallel to $\V{B}_0$, enabling propagating \Alfven waves to be
simulated.

The limitation to two spatial dimensions has two important effects: (i) it
precludes the possibility of initializing two perpendicularly
polarized, counterpropagating wavepackets (as initialized here), which
is the essential set up of the Parker-Moffatt problem to be studied;
and (ii) it significantly limits the nonlinear couplings that can
arise, constraining the results in an unphysical way
\citep{Howes:2014c}.

Although \citet{Pezzi:2017} state that they set up two \Alfvenic wavepackets, several pieces of evidence suggest that their initial wavepackets are indeed not \Alfvenic. First,
they initialize the fluctuations by imposing the constraint that
fluctuations are strictly transverse to the equilibrium magnetic field
$\V{B}_0 \cdot \delta \V{B}=0 $. Although this is a necessary
condition for an \Alfven wave, it is not sufficient: the polarization
of the magnetic field fluctuation for the \Alfven wave with wavevector
$\V{k}$ must also be oriented in the direction given by $\V{B}_0
\times \V{k}$ \citep{Maron:2001,Howes:2013a,Howes:2014c}. Any
component of the magnetic field fluctuation in the other direction of
the plane given by $\V{B}_0 \cdot \delta \V{B}=0$ represents not an
\Alfven wave, but a different mode: in incompressible MHD, it would
belong to the pseudo-\Alfven wave
\citep{Maron:2001,Howes:2013a,Howes:2014c}, and in compressible MHD,
it would represent some combination of the fast and slow magnetosonic
modes \citep{Cho:2003}. That their initialization in fact excites some
magnetosonic modes is supported by two results in their paper. First,
their initialization excites density fluctuations even before the
wavepackets collide, a feature they attribute to magnetic pressure
fluctuations due to finite-amplitude magnetic fluctuations, but that
could simply be because their initialization directly included these
linear magnetosonic modes.  Second, even before the waves collide in
the MHD simulation, shown in the upper left panel of their Figure~1,
one can see small disturbances well ahead of and behind each wavepacket,
suggesting that magnetosonic modes, which have a different group
velocity from the \Alfven mode, are dispersively spreading out the
wavepacket, a linear effect.

The second impact of the limitation to two spatial dimensions is that
the nonlinear couplings possible in the oblique 2D configuration are
unphysically constrained relative to the three dimensional case, as
explained in detail in \citet{Howes:2014c}.  This constraint can
artificially enhance the nonlinear coupling to magnetosonic modes
relative to the coupling to \Alfvenic modes, dramatically altering the
nonlinear evolution.

Our perspective is that any investigation of the so-called
Parker-Moffatt problem is most cleanly performed in three spatial
dimensions.  Our results show definitively that when two localized
\Alfven wavepackets collide, the interaction occurs only while they
overlap, ceasing when the wavepackets separate cleanly, leaving
distorted wavepackets that have essentially the same finite extent along
the equilibrium magnetic field.  Before concluding, it is also
worthwhile emphasizing that our results where obtained with a
gyrokinetic simulation that resolves all of the low-frequency---not
including cyclotron and fast magnetosonic physics
\citep{Howes:2006}---compressible and kinetic effects of a weakly
collisional plasma.

\section{Conclusion}
\label{sec:conc}
Our results here show that many of the fundamental properties of
\Alfven wave collisions, originally characterized analytically and
numerically in the idealized case of a periodic \Alfven wave
collision, persist under the more realistic conditions of a localized
\Alfven wavepacket collision. Specifically, we have demonstrated that
strong localized \Alfven wavepacket collisions are effective in
mediating the nonlinear cascade of energy to small perpendicular
scales, as demonstrated in \figref{fig:power}. Furthermore, we have
shown that strong localized \Alfven wavepacket collisions also
self-consistently generate current sheets that persist even between
collisions, as shown in Figures~\ref{fig:zcol} and~\ref{fig:zba},
confirming a robust mechanism to explain the ubiquitous current sheets
observed in numerical simulations of plasma turbulence
\citep{Wan:2012,Karimabadi:2013,TenBarge:2013a,Wu:2013,Zhdankin:2013}
and inferred from measurements of solar wind turbulence
\citep{Osman:2011,Borovsky:2011,Osman:2012a,Perri:2012a,Wang:2013,Wu:2013,Osman:2014b}.

The overall evolution of localized \Alfven wavepacket collisions
involves strong nonlinear interactions occurring while the wavepackets
overlap, followed by a clean separation of the wavepackets with
straight uniform magnetic fields in between and the cessation of
nonlinear evolution, as visualized in \figref{fig:3d}.  The
wavepackets remain localized along the equilibrium magnetic field
before and after each wavepacket collision. It is important to
emphasize that these characteristics are predicted based on the
properties of the incompressible MHD equations, but are found even in
the gyrokinetic simulation presented here which resolves dispersive
and kinetic effects beyond the reach of the MHD theory.  That these
important properties of \Alfven wave collisions persist in the
realistic localized wavepacket case further supports the contention
that \Alfven wave collisions represent the fundamental building block
of astrophysical plasma turbulence.

Previous analytical and numerical work in
the limit of weakly nonlinear, periodic \Alfven wave collisions,
\citep{Howes:2013a,Nielson:2013a} has shown that a nonlinearly
generated mode, which is not a solution of the linear dispersion
relation, serves to mediate the energy transfer to small perpendicular
scales. Future work will address the question of whether this
nonlinearly generated mode still plays a key role in the more
realistic case of localized \Alfven wavepacket collisions
\citep{Verniero:2017b}.


 This material is based upon work supported by the National Science Foundation Graduate Research Fellowship Program under Grant No. 1048957, NSF PHY-10033446, NSF CAREER AGS-1054061, and DOE
DE-SC0014599.  This research used resources of the Oak Ridge
Leadership Computing Facility, which is a DOE Office of Science User
Facility supported under Contract DE-AC05-00OR22725. We especially
thank one of their staff members, Dave Pugmire, for invaluable
assistance via email. This work used the Extreme Science and
Engineering Discovery Environment (XSEDE), which is supported by
National Science Foundation grant number ACI-1053575, through NSF
XSEDE Award PHY090084.

\appendix

\section{Windowing Function for Localized  \Alfven Wavepacket Collisions}
\label{appendix:setup}

\subsection{Parallel Dependence of \Alfven Eigenfunctions}

Initializing \Alfven wavepackets localized along the equilibrium
magnetic field using the \T{AstroGK} code \citep{Numata:2010} takes
advantage of the remarkable feature of gyrokinetics that the linear
collisionless gyrokinetic dispersion relation, when suitably
normalized, is independent of the parallel wavenumber of fluctuations,
as explained below.

For a fully ionized proton-electron plasma with Maxwellian equilibrium
velocity distributions and zero net drift velocities, the complex
frequency that is the solution of the Vlasov-Maxwell linear dispersion
relation \citep{Stix:1992,Quataert:1998} can be expressed as a
function of five dimensionless parameters, $\omega/\Omega_i =
\tilde{\omega}_{VM}(k_\parallel \rho_i, k_\perp \rho_i, \beta_i,
T_i/T_e, v_{ti}/c)$ \citep{Howes:2006}.  In the gyrokinetic limit of
non-relativistic ($v_{ti}/c\ll 1$), low-frequency ($\omega/\Omega_i
\ll 1$), anisotropic ($k_\parallel/k_\perp \ll 1$) fluctuations, the
complex eigenfrequency of the linear collisionless gyrokinetic
dispersion relation has just three dimensionless parameters when
normalized by $k_\parallel$, $\omega/(k_\parallel v_A) =
\overline{\omega}_{GK} (k_\perp \rho_i, \beta_i, T_i/T_e)$
\citep{Howes:2006}. The fact that the suitably normalized complex
eigenfrequency and complex Fourier coefficients of the eigenfunction
in gyrokinetics is independent of the parallel wavenumber $k_\parallel$
means that the phase and amplitude relationships among the different
components of the eigenfunction are independent of the parallel
wavenumber or coordinate.

Consider, for example, the procedure for initializing an \Alfven
wavepacket localized along the direction parallel to the equilibrium
magnetic field $\V{B}_0=B_0 \zhat$.  For a perpendicular Fourier
component of that wavepacket with a wavevector $(k_x,k_y)$, we specify
the complex perpendicular Fourier coefficient of the $y$ component of
the perturbed magnetic field by $\delta \hat{B}_y(k_x,k_y,z)$.  To
obtain the complex perpendicular Fourier coefficient of any other
component of the eigenfunction, for example the parallel magnetic
field perturbation, $\delta \hat{B}_z(k_x,k_y,z)$, we simply multiply the
ratio of the suitably normalized linear eigenfunction components,
\begin{equation}
  \delta \hat{B}_z(k_x,k_y,z) =\left[
  \frac{\delta\hat{\overline{B}}_z(k_x,k_y, \beta_i, T_i/T_e)}
       {\delta\hat{\overline{B}}_y(k_x,k_y, \beta_i, T_i/T_e)}\right]
  \delta \hat{B}_y(k_x,k_y,z)
  \label{eq:eigfact}
\end{equation}
The key property of gyrokinetics that we exploit is the fact that the
ratio in brackets, determined from the linear eigenfunction, is
independent of the parallel wavenumber $k_z$ and parallel coordinate
$z$.  For simplicity, we suppress the dependence in the plasma
parameters $(\beta_i, T_i/T_e)$ below, as they are constants for a
given calculation.

To demonstrate that the linear eigenfunction derived for a given
perpendicular wavevector $(k_x,k_y)$ can be used to construct a
wavepacket of arbitrary functional form in $z$, we note that the
Fourier transform in $z$ of a component of the fluctuating magnetic
field can be expressed as
\begin{equation}
  \delta \hat{B}_j(k_x,k_y,z) =\sum_{k_z}   \delta \hat{B}_j(k_x,k_y,k_z) e^{i k_zz},
  \label{eq:dbft}
\end{equation}
where the index $j$ denotes any component $x$, $y$, or $z$. By
substituting \eqref{eq:dbft} into the left side of \eqref{eq:eigfact}
for $ \delta \hat{B}_z(k_x,k_y,z) $ and right side of
\eqref{eq:eigfact} for $ \delta \hat{B}_y(k_x,k_y,z) $, we obtain
\begin{equation}
\sum_{k_z} \delta \hat{B}_z(k_x,k_y,k_z) e^{i k_zz}= \sum_{k_z} \left[
  \frac{\delta\hat{\overline{B}}_z(k_x,k_y)}
       {\delta\hat{\overline{B}}_y(k_x,k_y)}\right]
\delta \hat{B}_y(k_x,k_y,k_z) e^{i k_zz},
\end{equation}
where we have exploited the fact that the factor in brackets is independent of $k_z$
to bring it inside of the summation on the right-hand side. The orthogonality
of basis functions $e^{i k_zz}$ means that  the corresponding  terms in the sum must be equal, yielding the result
\begin{equation}
  \delta \hat{B}_j(k_x,k_y,k_z) =\left[
  \frac{\delta\hat{\overline{B}}_z(k_x,k_y, \beta_i, T_i/T_e)}
       {\delta\hat{\overline{B}}_y(k_x,k_y, \beta_i, T_i/T_e)}\right]
  \delta \hat{B}_y(k_x,k_y,k_z).
  \label{eq:dbftkz}
\end{equation}
Therefore, the phase and amplitude relationships between different
components of the eigenfunction are the same whether the parallel
dependence is expressed in coordinate space $z$ or Fourier space
$k_z$.

The procedure for initializing a localized wavepacket in \T{AstroGK}
with an arbitrary functional form in the parallel direction $z$
therefore follows: (i) the form of the wavepacket in the parallel
direction $z$ for the perpendicular Fourier component of the parallel
vector potential $\delta \hat{A}_\parallel(k_x,k_y,z)$ is specified;
(ii) the linear collisionless gyrokinetic dispersion relation
$\overline{\omega}_{GK} (k_\perp \rho_i, \beta_i, T_i/T_e)$ is solved
for the other suitably normalized electromagnetic field potentials
$\hat{\overline{\phi}}(k_x,k_y)$ and $\delta
\hat{\overline{B}}_\parallel(k_x,k_y)$ as well as the perturbed ion
and electron gyrokinetic distribution functions,
$\hat{\overline{h}}_i(v_\parallel, v_\perp,k_x,k_y)$ and
$\hat{\overline{h}}_e(v_\parallel, v_\perp,k_x,k_y)$; and (iii) the
ratio of the suitably normalized eigenfunctions are used to compute
the form of the wavepacket in the parallel direction $z$ for the
perpendicular Fourier component of all the other components of the
eigenfunction, for example $\delta \hat{B}_\parallel(k_x,k_y,z)$. This
completes the specification of a wavepacket of arbitrary parallel
functional dependence for the chosen linear gyrokinetic wave mode.

\subsection{Specification of Windowing Function}
To localize the \Alfven wavepacket, we use a windowing function $w(z)$
for the complex Fourier wave amplitude of each perpendicular Fourier
mode. The windowing function in $z$ for a given perpendicular Fourier
mode is
\begin{equation}
  w(z) = \exp \left[\left(\frac{z-z_0}{\Delta_z}\right)^p \right]
  + \exp \left[\left(\frac{z-z_0+L_z}{\Delta_z}\right)^p \right]
  + \exp \left[\left(\frac{z-z_0-L_z}{\Delta_z}\right)^p \right]
\end{equation}
where the three window function parameters are the center of the
window $z_0$, the width of the window $\Delta_z$ and the power $p$ of
the exponential.  Note that the second and third terms are needed to
ensure periodicity in $z$ when the window does not fall to zero at the
$z$ limits of the domain (it is important, since only two terms are
included, that the window does fall to zero by at least an additional
distance $L_z$ beyond each $z$ boundary).

The sinusoidal variation in $z$ of each perpendicular Fourier mode is given by
\begin{equation}
f(z)= \cos ( k_z z + \delta),
\end{equation}
where $k_z$ is the wavenumber along the equilibrium magnetic field and
$\delta$ is a phase-shift for that mode.  The total initial waveform
along $z$ is then given by the product of the sinusoidal mode and the
windowing function, $w(z)f(z)$.

For the localized \Alfven wavepacket collision simulation presented in
this paper, the parameters for the dipolar wave
(which propagates in the $-z$ direction) are $k_za_0=-3$, $\delta=0$
$z_0=\pi/2 a_0= L_\parallel/4$, $\Delta_z = 1/2 a_0$, and the default
power $p=2$. For the unipolar wave (which propagates in the $+z$
direction), the parameters are $k_za_0=+1$, $\delta=\pi/2$ $z_0=-\pi/2
a_0= -L_\parallel/4$, $\Delta_z = 1/2 a_0$, and the default power
$p=2$. Note that the non-zero phase $\delta$ for this unipolar
wavepacket means it has an initial form $- \sin (k_zz)$.

Note that the aspect ratio of the characteristic perpendicular length
scale to the characteristic parallel length scale $a_0/\rho_i \equiv
\epsilon \ll 1$ is the small gyrokinetic expansion parameter.  The
parallel domain scale is given by $k_\parallel a_0 =1$, where the
parallel domain length $L_\parallel = 2 \pi/k_\parallel$, so the
domain length can also be expressed as $L_\parallel = 2 \pi a_0$.  All
parallel length scales in \T{AstroGK} are normalized by $a_0$, whereas
all perpendicular length scales are normalized by $\rho_i$, or
equivalently dimensionless perpendicular wavenumber $k_\perp \rho_i$.


\begin{thebibliography}{47}
\expandafter\ifx\csname natexlab\endcsname\relax\def\natexlab#1{#1}\fi

\bibitem[{Abel} {\em et~al.\/}(2008){Abel}, {Barnes}, {Cowley}, {Dorland} \&
  {Schekochihin}]{Abel:2008}
{\sc {Abel}, I.~G., {Barnes}, M., {Cowley}, S.~C., {Dorland}, W. \&
  {Schekochihin}, A.~A.} 2008 {Linearized model Fokker-Planck collision
  operators for gyrokinetic simulations. I. Theory}. {\em Phys.~Plasmas\/} {\bf
  15}~(12), 122509.

\bibitem[{Barnes} {\em et~al.\/}(2009){Barnes}, {Abel}, {Dorland}, {Ernst},
  {Hammett}, {Ricci}, {Rogers}, {Schekochihin} \& {Tatsuno}]{Barnes:2009}
{\sc {Barnes}, M., {Abel}, I.~G., {Dorland}, W., {Ernst}, D.~R., {Hammett},
  G.~W., {Ricci}, P., {Rogers}, B.~N., {Schekochihin}, A.~A. \& {Tatsuno}, T.}
  2009 {Linearized model Fokker-Planck collision operators for gyrokinetic
  simulations. II. Numerical implementation and tests}. {\em Phys.~Plasmas\/}
  {\bf 16}~(7), 072107.

\bibitem[{Boldyrev} {\em et~al.\/}(2011){Boldyrev}, {Perez}, {Borovsky} \&
  {Podesta}]{Boldyrev:2011}
{\sc {Boldyrev}, S., {Perez}, J.~C., {Borovsky}, J.~E. \& {Podesta}, J.~J.}
  2011 {Spectral Scaling Laws in Magnetohydrodynamic Turbulence Simulations and
  in the Solar Wind}. {\em Astrophys.~J.~Lett.\/} {\bf 741}, L19.

\bibitem[{Borovsky}(2008)]{Borovsky:2008}
{\sc {Borovsky}, J.~E.} 2008 {Flux tube texture of the solar wind: Strands of
  the magnetic carpet at 1 AU?} {\em J.~Geophys.~Res.\/} {\bf 113}, A08110.

\bibitem[{Borovsky}(2010)]{Borovsky:2010}
{\sc {Borovsky}, J.~E.} 2010 {Contribution of Strong Discontinuities to the
  Power Spectrum of the Solar Wind}. {\em Phys.~Rev.~Lett.\/} {\bf 105}~(11),
  111102.

\bibitem[{Borovsky} \& {Denton}(2011)]{Borovsky:2011}
{\sc {Borovsky}, J.~E. \& {Denton}, M.~H.} 2011 {No Evidence for Heating of the
  Solar Wind at Strong Current Sheets}. {\em Astrophys.~J.~Lett.\/} {\bf 739},
  L61.

\bibitem[{Cho} \& {Lazarian}(2003)]{Cho:2003}
{\sc {Cho}, J. \& {Lazarian}, A.} 2003 {Compressible magnetohydrodynamic
  turbulence: mode coupling, scaling relations, anisotropy, viscosity-damped
  regime and astrophysical implications}. {\em Mon.~Not.~Roy.~Astron.~Soc.\/}
  {\bf 345}, 325--339.

\bibitem[{Drake} {\em et~al.\/}(2016){Drake}, {Howes}, {Rhudy}, {Terry},
  {Carter}, {Kletzing}, {Schroeder} \& {Skiff}]{Drake:2016}
{\sc {Drake}, D.~J., {Howes}, G.~G., {Rhudy}, J.~D., {Terry}, S.~K., {Carter},
  T.~A., {Kletzing}, C.~A., {Schroeder}, J.~W.~R. \& {Skiff}, F.} 2016
  {Measurements of the nonlinear beat wave produced by the interaction of
  counterpropagating Alfven waves}. {\em Phys.~Plasmas\/} {\bf 23}~(2), 022305.

\bibitem[{Drake} {\em et~al.\/}(2013){Drake}, {Schroeder}, {Howes}, {Kletzing},
  {Skiff}, {Carter} \& {Auerbach}]{Drake:2013}
{\sc {Drake}, D.~J., {Schroeder}, J.~W.~R., {Howes}, G.~G., {Kletzing}, C.~A.,
  {Skiff}, F., {Carter}, T.~A. \& {Auerbach}, D.~W.} 2013 {Alfv{\'e}n wave
  collisions, the fundamental building block of plasma turbulence. IV.
  Laboratory experiment}. {\em Physics of Plasmas\/} {\bf 20}~(7), 072901.

\bibitem[{Drake} {\em et~al.\/}(2014){Drake}, {Schroeder}, {Shanken}, {Howes},
  {Skiff}, {Kletzing}, {Carter} \& {Dorfman}]{Drake:2014}
{\sc {Drake}, D.~J., {Schroeder}, J.~W.~R., {Shanken}, B.~C., {Howes}, G.~G.,
  {Skiff}, F., {Kletzing}, C.~A., {Carter}, T.~A. \& {Dorfman}, S.} 2014
  {Analysis of Magnetic Fields in Inertial Alfv{\'e}n Wave Collisions}. {\em
  IEEE Trans. Plasma Sci.\/} {\bf 42}, 2534--2535.

\bibitem[{Frieman} \& {Chen}(1982)]{Frieman:1982}
{\sc {Frieman}, E.~A. \& {Chen}, L.} 1982 {Nonlinear gyrokinetic equations for
  low-frequency electromagnetic waves in general plasma equilibria}. {\em
  Phys.~Fluids\/} {\bf 25}, 502--508.

\bibitem[{Galtier} {\em et~al.\/}(2000){Galtier}, {Nazarenko}, {Newell} \&
  {Pouquet}]{Galtier:2000}
{\sc {Galtier}, S., {Nazarenko}, S.~V., {Newell}, A.~C. \& {Pouquet}, A.} 2000
  {A weak turbulence theory for incompressible magnetohydrodynamics}. {\em
  J.~Plasma Phys.\/} {\bf 63}, 447--488.

\bibitem[Goldreich \& Sridhar(1995)]{Goldreich:1995}
{\sc Goldreich, P. \& Sridhar, S.} 1995 {Toward a Theery of Interstellar
  Turbulence II. Strong Alfv\'enic Turbulence}. {\em Astrophys.~J.\/} {\bf
  438}, 763--775.

\bibitem[{Howes}(2014)]{Howes:2014c}
{\sc {Howes}, G.~G.} 2014 The inherently three-dimensional nature of magnetized
  plasma turbulence. {\em Journal of Plasma Physics\/} {\bf FirstView}, 1--19.

\bibitem[{Howes}(2016)]{Howes:2016b}
{\sc {Howes}, G.~G.} 2016 {The Dynamical Generation of Current Sheets in
  Astrophysical Plasma Turbulence}. {\em Astrophys.~J.~Lett.\/} {\bf 827}, L28.

\bibitem[{Howes} \& {Bourouaine}(2017)]{Howes:2017b}
{\sc {Howes}, G.~G. \& {Bourouaine}, S.} 2017 {The Development of Magnetic
  Field Line Wander by Plasma Turbulence}. {\em J.~Plasma Phys.\/} Submitted.

\bibitem[{Howes} {\em et~al.\/}(2006){Howes}, {Cowley}, {Dorland}, {Hammett},
  {Quataert} \& {Schekochihin}]{Howes:2006}
{\sc {Howes}, G.~G., {Cowley}, S.~C., {Dorland}, W., {Hammett}, G.~W.,
  {Quataert}, E. \& {Schekochihin}, A.~A.} 2006 {Astrophysical Gyrokinetics:
  Basic Equations and Linear Theory}. {\em Astrophys.~J.\/} {\bf 651},
  590--614.

\bibitem[{Howes} {\em et~al.\/}(2012){Howes}, {Drake}, {Nielson}, {Carter},
  {Kletzing} \& {Skiff}]{Howes:2012b}
{\sc {Howes}, G.~G., {Drake}, D.~J., {Nielson}, K.~D., {Carter}, T.~A.,
  {Kletzing}, C.~A. \& {Skiff}, F.} 2012 {Toward Astrophysical Turbulence in
  the Laboratory}. {\em Phys.~Rev.~Lett.\/} {\bf 109}~(25), 255001.

\bibitem[{Howes} \& {Nielson}(2013)]{Howes:2013a}
{\sc {Howes}, G.~G. \& {Nielson}, K.~D.} 2013 {Alfv{\'e}n wave collisions, the
  fundamental building block of plasma turbulence. I. Asymptotic solution}.
  {\em Phys.~Plasmas\/} {\bf 20}~(7), 072302.

\bibitem[{Howes} {\em et~al.\/}(2013){Howes}, {Nielson}, {Drake}, {Schroeder},
  {Skiff}, {Kletzing} \& {Carter}]{Howes:2013b}
{\sc {Howes}, G.~G., {Nielson}, K.~D., {Drake}, D.~J., {Schroeder}, J.~W.~R.,
  {Skiff}, F., {Kletzing}, C.~A. \& {Carter}, T.~A.} 2013 {Alfv{\'e}n wave
  collisions, the fundamental building block of plasma turbulence. III. Theory
  for experimental design}. {\em Physics of Plasmas\/} {\bf 20}~(7), 072304.

\bibitem[Iroshnikov(1963)]{Iroshnikov:1963}
{\sc Iroshnikov, R.~S.} 1963 The turbulence of a conducting fluid in a strong
  magnetic field. {\em Astron. Zh.\/} {\bf 40}, 742, {English} Translation:
  Sov. Astron., 7 566 (1964).

\bibitem[{Karimabadi} {\em et~al.\/}(2013){Karimabadi}, {Roytershteyn}, {Wan},
  {Matthaeus}, {Daughton}, {Wu}, {Shay}, {Loring}, {Borovsky}, {Leonardis},
  {Chapman} \& {Nakamura}]{Karimabadi:2013}
{\sc {Karimabadi}, H., {Roytershteyn}, V., {Wan}, M., {Matthaeus}, W.~H.,
  {Daughton}, W., {Wu}, P., {Shay}, M., {Loring}, B., {Borovsky}, J.,
  {Leonardis}, E., {Chapman}, S.~C. \& {Nakamura}, T.~K.~M.} 2013 {Coherent
  structures, intermittent turbulence, and dissipation in high-temperature
  plasmas}. {\em Phys.~Plasmas\/} {\bf 20}~(1), 012303.

\bibitem[Kraichnan(1965)]{Kraichnan:1965}
{\sc Kraichnan, R.~H.} 1965 Inertial range spectrum of hyromagnetic turbulence.
  {\em Phys.~Fluids\/} {\bf 8}, 1385--1387.

\bibitem[Maron \& Goldreich(2001)]{Maron:2001}
{\sc Maron, J. \& Goldreich, P.} 2001 Simulations of incompressible
  magnetohydrodynamic turbulence. {\em Astrophys.~J.\/} {\bf 554}, 1175--1196.

\bibitem[{Matthaeus} \& {Montgomery}(1980)]{Matthaeus:1980}
{\sc {Matthaeus}, W.~H. \& {Montgomery}, D.} 1980 {Selective decay hypothesis
  at high mechanical and magnetic Reynolds numbers}. {\em Annals of the New
  York Academy of Sciences\/} {\bf 357}, 203--222.

\bibitem[{Meneguzzi} {\em et~al.\/}(1981){Meneguzzi}, {Frisch} \&
  {Pouquet}]{Meneguzzi:1981}
{\sc {Meneguzzi}, M., {Frisch}, U. \& {Pouquet}, A.} 1981 {Helical and
  nonhelical turbulent dynamos}. {\em Phys.~Rev.~Lett.\/} {\bf 47}, 1060--1064.

\bibitem[{Montgomery} \& {Matthaeus}(1995)]{Montgomery:1995}
{\sc {Montgomery}, D. \& {Matthaeus}, W.~H.} 1995 {Anisotropic Modal Energy
  Transfer in Interstellar Turbulence}. {\em Astrophys.~J.\/} {\bf 447}, 706.

\bibitem[{Ng} \& {Bhattacharjee}(1996)]{Ng:1996}
{\sc {Ng}, C.~S. \& {Bhattacharjee}, A.} 1996 {Interaction of Shear-Alfven Wave
  Packets: Implication for Weak Magnetohydrodynamic Turbulence in Astrophysical
  Plasmas}. {\em Astrophys.~J.\/} {\bf 465}, 845.

\bibitem[{Nielson} {\em et~al.\/}(2013){Nielson}, {Howes} \&
  {Dorland}]{Nielson:2013a}
{\sc {Nielson}, K.~D., {Howes}, G.~G. \& {Dorland}, W.} 2013 {Alfv{\'e}n wave
  collisions, the fundamental building block of plasma turbulence. II.
  Numerical solution}. {\em Physics of Plasmas\/} {\bf 20}~(7), 072303.

\bibitem[{Numata} {\em et~al.\/}(2010){Numata}, {Howes}, {Tatsuno}, {Barnes} \&
  {Dorland}]{Numata:2010}
{\sc {Numata}, R., {Howes}, G.~G., {Tatsuno}, T., {Barnes}, M. \& {Dorland},
  W.} 2010 {AstroGK: Astrophysical gyrokinetics code}. {\em J.~Comp.~Phys.\/}
  {\bf 229}, 9347.

\bibitem[{Osman} {\em et~al.\/}(2014){Osman}, {Matthaeus}, {Gosling}, {Greco},
  {Servidio}, {Hnat}, {Chapman} \& {Phan}]{Osman:2014b}
{\sc {Osman}, K.~T., {Matthaeus}, W.~H., {Gosling}, J.~T., {Greco}, A.,
  {Servidio}, S., {Hnat}, B., {Chapman}, S.~C. \& {Phan}, T.~D.} 2014 {Magnetic
  Reconnection and Intermittent Turbulence in the Solar Wind}. {\em
  Phys.~Rev.~Lett.\/} {\bf 112}~(21), 215002.

\bibitem[{Osman} {\em et~al.\/}(2011){Osman}, {Matthaeus}, {Greco} \&
  {Servidio}]{Osman:2011}
{\sc {Osman}, K.~T., {Matthaeus}, W.~H., {Greco}, A. \& {Servidio}, S.} 2011
  {Evidence for Inhomogeneous Heating in the Solar Wind}. {\em
  Astrophys.~J.~Lett.\/} {\bf 727}, L11.

\bibitem[{Osman} {\em et~al.\/}(2012){Osman}, {Matthaeus}, {Wan} \&
  {Rappazzo}]{Osman:2012a}
{\sc {Osman}, K.~T., {Matthaeus}, W.~H., {Wan}, M. \& {Rappazzo}, A.~F.} 2012
  {Intermittency and Local Heating in the Solar Wind}. {\em Phys.~Rev.~Lett.\/}
  {\bf 108}~(26), 261102.

\bibitem[{Perri} {\em et~al.\/}(2012){Perri}, {Goldstein}, {Dorelli} \&
  {Sahraoui}]{Perri:2012a}
{\sc {Perri}, S., {Goldstein}, M.~L., {Dorelli}, J.~C. \& {Sahraoui}, F.} 2012
  {Detection of Small-Scale Structures in the Dissipation Regime of Solar-Wind
  Turbulence}. {\em Phys.~Rev.~Lett.\/} {\bf 109}~(19), 191101.

\bibitem[{Pezzi} {\em et~al.\/}(2017){Pezzi}, {Parashar}, {Servidio},
  {Valentini}, {V{\'a}sconez}, {Yang}, {Malara}, {Matthaeus} \&
  {Veltri}]{Pezzi:2017}
{\sc {Pezzi}, O., {Parashar}, T.~N., {Servidio}, S., {Valentini}, F.,
  {V{\'a}sconez}, C.~L., {Yang}, Y., {Malara}, F., {Matthaeus}, W.~H. \&
  {Veltri}, P.} 2017 {Revisiting a Classic: The Parker-Moffatt Problem}. {\em
  Astrophys.~J.\/} {\bf 834}, 166.

\bibitem[{Quataert}(1998)]{Quataert:1998}
{\sc {Quataert}, E.} 1998 {Particle Heating by Alfv\'enic Turbulence in Hot
  Accretion Flows}. {\em Astrophys.~J.\/} {\bf 500}, 978--991.

\bibitem[{Sridhar} \& {Goldreich}(1994)]{Sridhar:1994}
{\sc {Sridhar}, S. \& {Goldreich}, P.} 1994 {Toward a theory of interstellar
  turbulence. 1: Weak Alfvenic turbulence}. {\em Astrophys.~J.\/} {\bf 432},
  612--621.

\bibitem[{Stix}(1992)]{Stix:1992}
{\sc {Stix}, T.~H.} 1992 {\em {Waves in Plasmas}\/}. New York: American
  Institute of Physics.

\bibitem[{TenBarge} \& {Howes}(2013)]{TenBarge:2013a}
{\sc {TenBarge}, J.~M. \& {Howes}, G.~G.} 2013 {Current Sheets and
  Collisionless Damping in Kinetic Plasma Turbulence}. {\em
  Astrophys.~J.~Lett.\/} {\bf 771}, L27.

\bibitem[{Uritsky} {\em et~al.\/}(2010){Uritsky}, {Pouquet}, {Rosenberg},
  {Mininni} \& {Donovan}]{Uritsky:2010}
{\sc {Uritsky}, V.~M., {Pouquet}, A., {Rosenberg}, D., {Mininni}, P.~D. \&
  {Donovan}, E.~F.} 2010 {Structures in magnetohydrodynamic turbulence:
  Detection and scaling}. {\em Phys.~Rev.~E\/} {\bf 82}~(5), 056326.

\bibitem[{Valentini} {\em et~al.\/}(2007){Valentini}, {Tr{\'a}vn{\'{\i}}{\v
  c}ek}, {Califano}, {Hellinger} \& {Mangeney}]{Valentini:2007}
{\sc {Valentini}, F., {Tr{\'a}vn{\'{\i}}{\v c}ek}, P., {Califano}, F.,
  {Hellinger}, P. \& {Mangeney}, A.} 2007 {A hybrid-Vlasov model based on the
  current advance method for the simulation of collisionless magnetized
  plasma}. {\em J.~Comp.~Phys.\/} {\bf 225}, 753--770.

\bibitem[{Verniero} \& {Howes}(2017)]{Verniero:2017b}
{\sc {Verniero}, J.~L. \& {Howes}, G.~G.} 2017 {The Physics of Energy Transfer
  in \Alfven wave collisions: Periodic Plane Waves vs.~Isolated Wavepackets}.
  {\em J.~Plasma Phys.\/} In preparation.

\bibitem[{Wan} {\em et~al.\/}(2012){Wan}, {Matthaeus}, {Karimabadi},
  {Roytershteyn}, {Shay}, {Wu}, {Daughton}, {Loring} \& {Chapman}]{Wan:2012}
{\sc {Wan}, M., {Matthaeus}, W.~H., {Karimabadi}, H., {Roytershteyn}, V.,
  {Shay}, M., {Wu}, P., {Daughton}, W., {Loring}, B. \& {Chapman}, S.~C.} 2012
  {Intermittent Dissipation at Kinetic Scales in Collisionless Plasma
  Turbulence}. {\em Phys.~Rev.~Lett.\/} {\bf 109}~(19), 195001.

\bibitem[{Wang} {\em et~al.\/}(2013){Wang}, {Tu}, {He}, {Marsch} \&
  {Wang}]{Wang:2013}
{\sc {Wang}, X., {Tu}, C., {He}, J., {Marsch}, E. \& {Wang}, L.} 2013 {On
  Intermittent Turbulence Heating of the Solar Wind: Differences between
  Tangential and Rotational Discontinuities}. {\em Astrophys.~J.~Lett.\/} {\bf
  772}, L14.

\bibitem[{Wu} {\em et~al.\/}(2013){Wu}, {Perri}, {Osman}, {Wan}, {Matthaeus},
  {Shay}, {Goldstein}, {Karimabadi} \& {Chapman}]{Wu:2013}
{\sc {Wu}, P., {Perri}, S., {Osman}, K., {Wan}, M., {Matthaeus}, W.~H., {Shay},
  M.~A., {Goldstein}, M.~L., {Karimabadi}, H. \& {Chapman}, S.} 2013
  {Intermittent Heating in Solar Wind and Kinetic Simulations}. {\em
  Astrophys.~J.~Lett.\/} {\bf 763}, L30.

\bibitem[{Zhdankin} {\em et~al.\/}(2012){Zhdankin}, {Boldyrev}, {Mason} \&
  {Perez}]{Zhdankin:2012}
{\sc {Zhdankin}, V., {Boldyrev}, S., {Mason}, J. \& {Perez}, J.~C.} 2012
  {Magnetic Discontinuities in Magnetohydrodynamic Turbulence and in the Solar
  Wind}. {\em Phys.~Rev.~Lett.\/} {\bf 108}~(17), 175004.

\bibitem[{Zhdankin} {\em et~al.\/}(2013){Zhdankin}, {Uzdensky}, {Perez} \&
  {Boldyrev}]{Zhdankin:2013}
{\sc {Zhdankin}, V., {Uzdensky}, D.~A., {Perez}, J.~C. \& {Boldyrev}, S.} 2013
  {Statistical Analysis of Current Sheets in Three-dimensional
  Magnetohydrodynamic Turbulence}. {\em Astrophys.~J.\/} {\bf 771}, 124.

\end{thebibliography}

\end{document}